\documentclass[nofootinbib,twocolumn,superscriptaddress]{revtex4-1}
\usepackage[usenames,dvipsnames]{color,xcolor}
\usepackage[english, frenchb]{babel}
\usepackage[latin1]{inputenc}
\usepackage{graphicx,amssymb}
\usepackage{hyperref}

\hypersetup{colorlinks, citecolor=Red, linkcolor=Blue, urlcolor=Red}

\begin{document}
\title{History-independence of steady state in simultaneous two-phase
  flow through porous media}

\author{Marion Erpelding}
\affiliation{Department of Physics, University of Oslo, PB 1048
Blindern, N-0316 Oslo, Norway}
\author{Santanu Sinha}
\affiliation{Department of Physics, Norwegian University of 
Science and Technology, N-7491 Trondheim, Norway}
\author{Ken Tore Tallakstad}
\affiliation{Department of Physics, University of Oslo, PB 1048
Blindern, N-0316 Oslo, Norway}
\author{Alex Hansen}
\affiliation{Department of Physics, Norwegian University of 
Science and Technology, N-7491 Trondheim, Norway}
\author{Eirik Grude Flekk{\o}y}
\affiliation{Department of Physics, University of Oslo, PB 1048
Blindern, N-0316 Oslo, Norway}
\author{Knut J{\o}rgen M{\aa}l{\o}y}
\affiliation{Department of Physics, University of Oslo, PB 1048
Blindern, N-0316 Oslo, Norway}

\date{\today}

\begin{abstract}
It is well known that the transient behavior during drainage or
imbibition in multiphase flow in porous media strongly depends on the history and initial condition of the system. However, when the steady-state regime is reached and both drainage and imbibition take place at
the pore level, the influence of the evolution history and initial preparation is an open question. Here, we
present an extensive experimental and numerical work investigating the history dependence of simultaneous steady-state two-phase flow
through porous media. Our experimental system consists of a Hele-Shaw
cell filled with glass beads which we model numerically by a network
of disordered pores transporting two immiscible fluids. From the
measurements of global pressure evolution, histogram of saturation and
cluster-size distributions, we find that when both phases are
flowing through the porous medium, the steady state does not depend on the
initial preparation of the system or on the way it has been
reached.
\end{abstract}

\maketitle

\section{Introduction}
Understanding the physical mechanisms underlying multiphase flow in porous media is crucial to a wide variety of industrial and environmental problems, such as oil recovery, $\text{CO}_2$ transport and storage or ground water management.
At the macroscopic scale, the flow of immiscible fluids in a porous
material is described by specifying the relations between global
quantities such as flow rate, pressure gradient or fluid
saturation. At the pore scale, this flow is governed by the
competition between capillary, viscous and gravitational
forces. Understanding the link between the two levels of description
requires to relate the position and shape of the interface(s) between
the two phases to the values of the macroscopic variables. From an
experimental point of view, two-dimensional model porous media
providing direct pore-scale visualization of the flow structures are
ideal tools to study multiphase flow mechanisms and their relations
with global quantities in controlled, laboratory-scale
situations. Over the past decades, micro channel networks etched in
transparent plates \cite{ap95a}, or prepared using molding techniques
\cite{bonnet77} and porous Hele-Shaw cells consisting of a layer of
beads between two parallel plates \cite{mfj85} have become classical
tools, to which improvements have been constantly proposed
\cite{fmsh97, berejnov08}. Experimental observations have been
explained through extensive numerical simulations based on network
models \cite{koplik85, lenormand88, blunt90, cp91, cp96} and lattice
Boltzmann methods \cite{rothman90, rothman91, rothman93, rothman95,
  liu12, aursjo2010, aursjo2011}, statistical models \cite{ws81, ww83, paterson84} and
differential equations \cite{binning99}. Most of the research in this
area has been focused upon transient phenomena, \textit{i.e.} drainage
or imbibition -- arising when one phase displaces the other in a
porous medium. The relation between macroscopic flow parameters, fluid
morphology and stability of the interface between the two phases has
been thoroughly observed \cite{mfj85,fmsh97} and successfully modeled
\cite{ww83,ws81}.

In this article, we deal with \textit{steady-state flow}, where many
questions are yet to be answered. In the steady-state regime, two phases are
injected simultaneously into the porous medium and one observes that
one or both are fragmented and transported in the form of clusters of various sizes, forming a complex flow pattern
with multiple interfaces. After a characteristic time, the system
reaches a steady-state in which the macroscopic flow variables
fluctuate around constant values. The usual distinction between
drainage and imbibition is irrelevant to describe steady-state
two-phase flow, in which both processes occur simultaneously. New
approaches are thus needed to understand this regime. In an effort to
bring new insight, experimental and numerical studies have
investigated the relations between macroscopic flow variables, and
different models have been proposed to relate them to pore-scale flow
mechanisms: Payatakes and co-workers carried out detailed
experimental, numerical and theoretical studies of steady-state
two-phase flow, with emphasis on the determination of relative
permeabilities \cite{ap95a, ap95b, ap99, tap07, cp91, cp96,
  vcp98}. The steady-state characteristics of macroscopic flow
properties have also been investigated numerically by Knudsen
\textit{et al.} \cite{kh02}. A power-law relation between
pressure and steady-state flow rate has been observed experimentally 
by Tallakstad \textit{et al.} in a 2D system \cite{tkrlmtf09, tallakstad09}, and by Rassi \textit{et al.} in a 3D system \cite{rassi2011}. Very
recently, the relation between the steady-state flow rate and pressure
drop has been derived analytically for two-phase flow through single
capillaries \cite{sinha13} and through porous media \cite{tkrlmtf09, tallakstad09,sinha12} and
 also supported by extensive numerical simulation. Distributions of
non-wetting or wetting clusters in the steady state have also been
studied experimentally by Tallakstad \textit{et al.} \cite{tkrlmtf09,
  tallakstad09} and numerically by Ramstad and Hansen \cite{rh06}, and
critical exponents were measured. It is worth noting though, that the
comparison of experimental and numerical results is not always
straightforward, due to the different boundary conditions used in the
two cases. Yet from the point of
view of statistical physics, the existence of a genuine steady state is very
promising to build a thermodynamic-like theoretical description of the
system. In that context, it is crucial to determine whether the
steady state is independent on the history of the process, or in other
words, whether it is a real \textit{state} in a thermodynamic sense
\cite{hr09}. It is well known that when drainage and imbibition occur
successively, the relative permeabilities become history dependent and
the pressure-saturation curves display an hysteresis \cite{juanes2006,
  aryana12}, the underlying pore-scale mechanisms of which are
known. Besides in the well-known magnetic and elastic systems, such
history dependence and hysteresis has also been observed in different flow
processes like hydrodynamic heat flow \cite{greytak73} and particle
flow through random media \cite{watson96}.  However, in the case of
two-phase flow through porous media, it is not trivial to predict
whether such an hysteresis will come into play in the steady-state
situation when drainage and imbibition occur simultaneously. It was
proposed that a thermodynamic-like description for simultaneous
two-phase flow in porous media can be sketched \cite{hr09} if the flow
was history independent.

Here, we present an extensive experimental and numerical study
in order to investigate the history-independence of the
steady state. Our experimental system consists of a porous Hele-shaw
cell in which we simultaneously inject air and a viscous
water-glycerol solution. We then compare steady-states obtained for a
given flow rate with different initial conditions. We model the system
by a network of disordered pores transporting two immiscible fluids. From pressure measurements and analysis of the statistical
properties of the flow patterns, we observe no history-dependence for
the steady state.

\section{Experimental setup}\label{section-setup}

\begin{figure}
\centering
\includegraphics[width=0.5\textwidth]{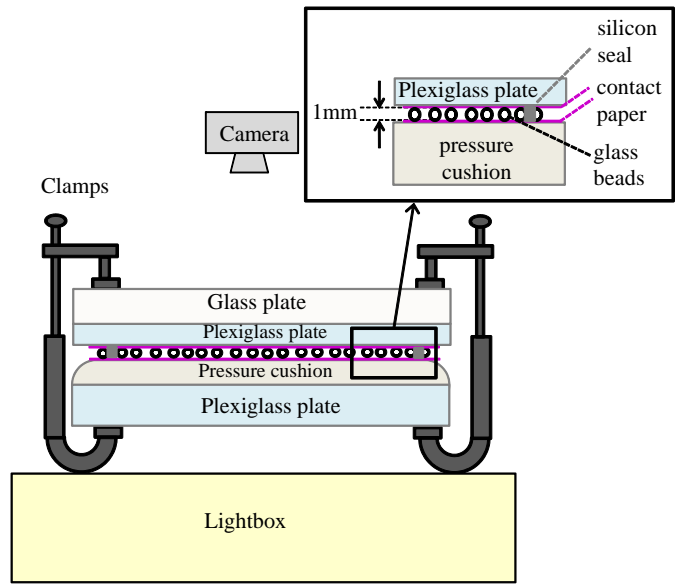}
\caption{Sketch of the experimental setup: the 2D porous matrix
  consists of a disordered mono layer of glass beads spread between
  two sheets of contact paper. The boundaries are sealed with silicon
  glue. The upper part of the system consists of a Plexiglas plate
  with drilled inlet and outlet flow channels. A pressure cushion and
  a thick glass plate placed below and above the porous matrix ensure
  the overall rigidity of the system and maintain its thickness
  constant. Clamps maintain all the layers together. A lightbox
  illuminates the system from below and a digital camera is placed
  above to record images of the flow structure.}\label{setupside}
\end{figure}
   
\begin{figure}
\centering
\includegraphics[width=0.5\textwidth]{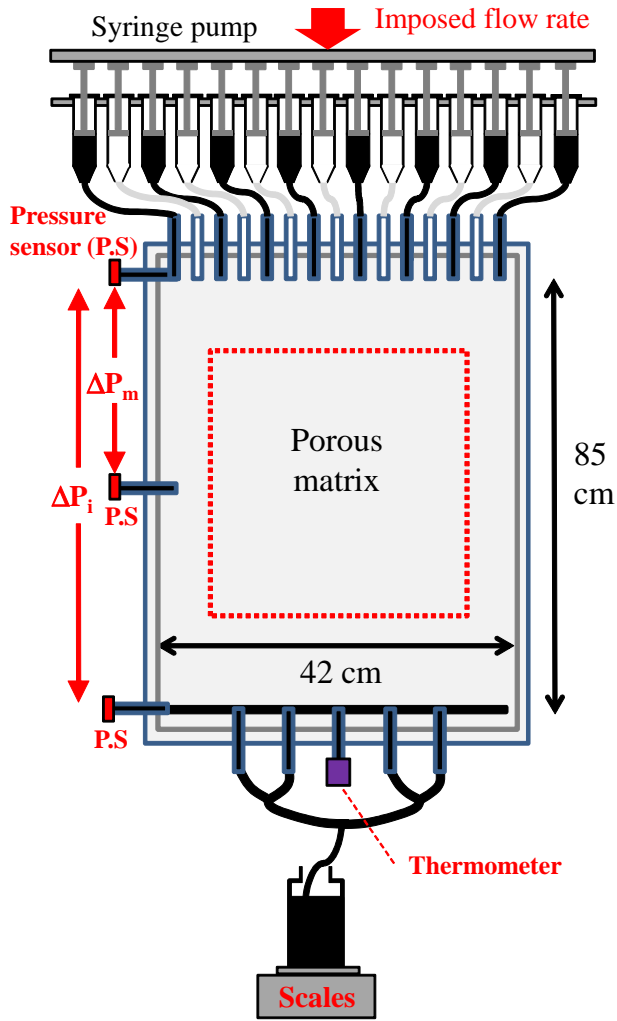}
\caption{Sketch of the experimental setup with the injection
  system. The two phases are contained in 15 syringes, each connected
  to one of the 15 inlet nodes of the porous model (7 syringes of air
  represented in white and 8 syringes of water-glycerol solution in
  black). The same syringe pump is used to inject both phases
  simultaneously. The dotted lines give the dimensions of the area
  studied by image analysis (note that proportions are not
  respected).}\label{setuptop}
\end{figure}

We use a two-dimensional, transparent, porous Hele-shaw cell. This
experimental setup, shown on Figures \ref{setupside} and
\ref{setuptop}, has been described in details in \cite{tallakstad09}
and we recall its main features here. The porous medium consists of a
random mono layer of glass beads, $1\text{ mm}$ in diameter, spread between
the sticky sides of two sheets of contact paper. Its lateral
boundaries are sealed with silicon glue. Attached on top of this
layer, a Plexiglas plate with etched flow channels allows
injection and evacuation of fluids into and from the porous matrix. A
pressure cushion (see \cite{tallakstad09} for details) placed below
the porous medium, and a thick glass plate on top prevent the system
from bending when the pressure increases as fluids are
injected. Clamps placed all around the setup maintain all the layers
together. This way, we obtain a porous medium of constant thickness
$a=1\text{ mm}$, length $L=85\text{ cm}$, width $W=42\text{ cm}$ in which the beads remain
immobile. The porosity $\phi$ and absolute permeability $\kappa_0$ of the
medium are found experimentally to be $\phi=0.63$ and
$\kappa_0=1.95\cdot10^{5}\text{ cm}^2$ \cite{tallakstad09}.

\begin{figure}
\centering
\includegraphics[width=0.5\textwidth]{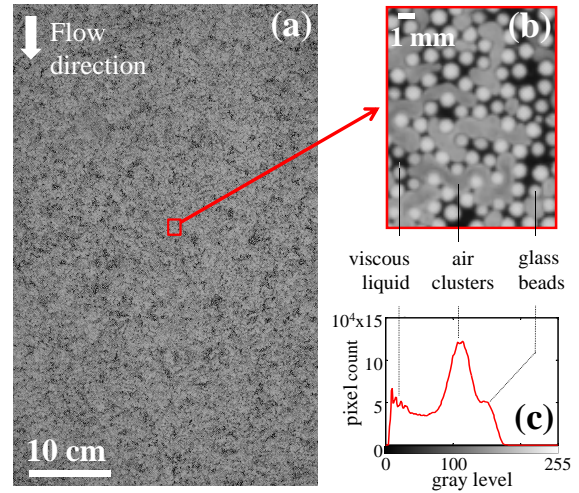}
\caption{(a) Example of steady-state image ($2400 \times 3800$
  pixels): the flow pattern is made of air clusters (gray) of various
  sizes surrounded by the viscous liquid (black) -- (b) Zoom: the high
  image resolution makes it possible to distinguish glass beads
  (bright gray), air clusters (gray) and viscous liquid (black). These
  three phases yield three ``peaks'' on the gray scale image
  histograms, as illustrated by (c). The height of the peaks contains
  information about the saturation of the system.} \label{flowimage}
\end{figure}

We use the same fluid pair as Tallakstad \textit{et al.}
\cite{tallakstad09}, namely air as the non-wetting phase and a viscous
water-glycerol solution (15-85\% in mass) as the wetting phase. The
latter is dyed in black with 0.1\% Negrosine to obtain a good contrast
on the experimental images (see Figure \ref{flowimage}).  As pointed
out by Tallakstad \textit{et al.}, the use of air as one of the phases
introduces more complexity in the two-phase flow problem for two
reasons: first, it yields a high viscosity contrast with the glycerol
solution, and second, its compressibility gives rise to rapid bursts
or avalanches \cite{tallakstad09}. However, from an experimental point
of view, it has the huge advantage of allowing us to reuse the same
porous model for all experiments. Indeed, it can easily be flushed
out, making it possible to obtain reproducible initial
conditions. Therefore, we find it fully convenient for the present
study. As shown on Figure \ref{setuptop}, the two phases are injected
simultaneously, using the same syringe pump, from 15 syringes  (7 of
air and 8 of water-glycerol solution), each connected to one of the 15
inlet nodes of the porous model. In the following, we note $Q$ the total flow rate, while $Q_{\rm w}=(8/15)Q$ and $Q_{\rm nw}=(7/15)Q$ denote the wetting and non-wetting flow rates, respectively. To account for small temperature
variations due to the heat released by the lightbox (see Figure
\ref{setupside}), we monitor the temperature of the wetting phase at
the outlet of the model. The viscosity $\mu_w$ of the wetting phase is
deduced accordingly using the empirical formula given in
\cite{Cheng2008}. In the series of experiments presented here, the
measured temperatures are in the range $24.4 - 29.3^\circ$ Celsius,
giving $0.083>\mu_w>0.062\text{ Pa.s}$. The viscosity of air being $\mu_{nw}\approx 1.9\times 10^{-5}\text{ Pa.s}$, the viscosity ratio $\rm M=\mu_{nw}/\mu_{w}$ of the order of $10^{-4}$ in all experiments.

With this setup, the total flow rate $Q$ and the
fractional flow $F_{\rm{w}}=Q_{\rm{w}}/Q$ are controlled flow variables. The volumes of wetting and non-wetting fluids present in the porous matrix, $V_{\rm w}$ and $V_{\rm nw}$, are free to vary with time. Thus, the saturations $S_{\rm w}=V_{\rm w}/V$ and  $S_{\rm nw}=V_{\rm nw}/V$, where $V$ is the total pore volume, are free to fluctuate. The flow
is characterized by the capillary number:
\begin{equation}
 {\rm Ca} =\frac{\mu_wQ_w}{\gamma A},
 \label{eq:Ca}
\end{equation}
where $\mu_w$ is the wetting phase viscosity, $Q_w$ is the total
wetting fluid flow rate, $\gamma\approx6.4\cdot10^{-2} \text{ N/m}$ is the interfacial
tension between the two phases \cite{tallakstad09} and $A=Wa\phi$ is the cross-section of the
porous matrix. In the present experiments, we have explored the range
$3.33\ 10^{-6}\leqslant \rm Ca \leqslant 1.13\ 10^{-4}$. The highest
experimental value of $\rm Ca$ is set by the maximum pressure that the
porous model can hold. However, as we will see in Section
\ref{section-network-model}, we have also explored higher values of
$\rm Ca$ in numerical simulations, namely $1.92\times 10^{-5}
\leqslant \rm Ca \leqslant 7.0\times 10^{-2}$. \\

Our analysis of steady-state flow and its history dependence relies on
two kinds of information: measurements of the pressure inside the
model, and pictures of the flow pattern. We measure the pressure
$P(t)$ in the wetting phase as a function of time $t$ using
flow-through pressure sensors (SensorTechnics 26PC0100G6G) placed at
three different points of the model as indicated on Figure
\ref{setuptop}. The porous model is lit from below using a lightbox,
and images of the flow structure are recorded regularly using a Nikon
D200 digital reflex camera giving $2592\times3872$ pixels images with
a spatial resolution of $8\times8 $ pixels per $\text{mm}^2$. Before
further processing, images are cropped to remove boundaries, leaving
us with $2400\times3800$ pixels on the final images. Figure
\ref{flowimage}(a) shows an example of steady-state image. The area of
the imaged zone (see Figure \ref{setuptop}) is large enough to contain
many air clusters of various sizes. As illustrated by Figure
\ref{flowimage}(b), the high image resolution makes it possible to
distinguish glass beads, air clusters, and viscous liquid. Each of
these phases gives rise to a peak on the gray-scale image histogram
(see Figure \ref{flowimage}(c)). The heights of these peaks contain
information about the proportion of wetting and non-wetting fluids in
the system, thus about the saturations $S_{\rm w}$ and $S_{\rm nw}$.

\section{Experimental results and discussion}

\subsection{Principle of the experiments}\label{sec-exp-procedure}

\begin{figure}[htp]
\centering
\includegraphics[width=0.45\textwidth]{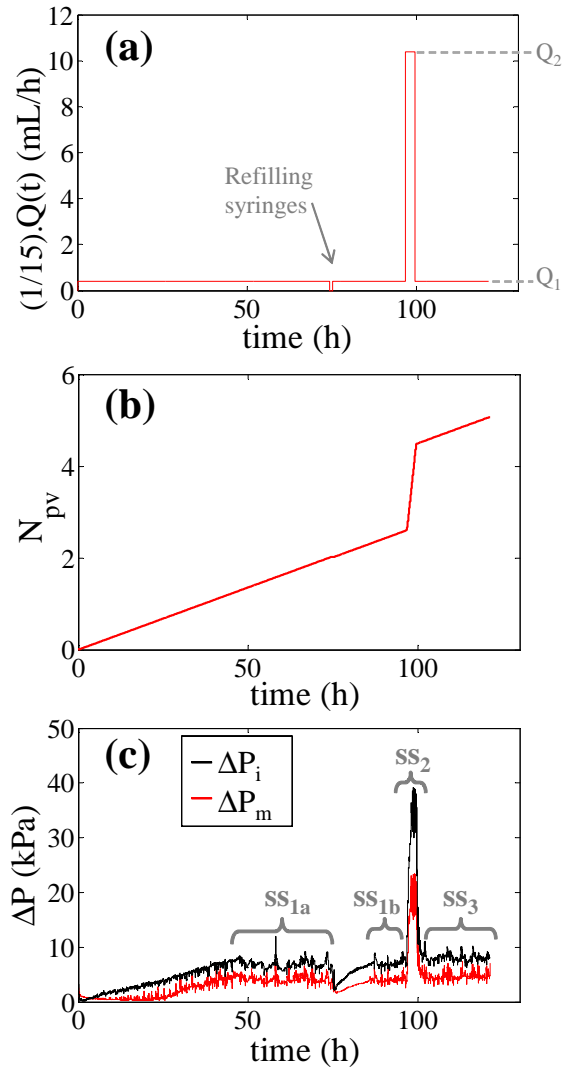}
\caption{Temporal variation of imposed and measured experimental
  variables for a typical experiment (Exp.3, see Table
  \ref{expparams}): (a) Temporal variation of the imposed flow rate
  $\left(1/15\right).Q$ for one syringe. -- (b) Ratio of the total injected fluid
  volume to the available pore volume, computed from the imposed flow
  rate using Eq. \ref{eq:Npv} -- (c) Corresponding pressure drop
  measurements. Plateaus characterize the different steady-states
  ($ss$). The sudden drop around $t=75h$ is an experimental artifact
  due to the refilling of the syringes, as explained in the
  text. Steady-states obtained before and after refilling are
  distinguished using the subscripts $a$ and $b$.}\label{pressresults}
\end{figure}

We investigate the history-dependence of steady-state flow by
comparing steady states obtained at the same flow rate but with
different initial conditions. For this, we have performed experiments
in which the flow rate is modified twice, as illustrated by Figure
\ref{pressresults}(a). The porous model is initially filled with the
wetting phase only. Then, both phases are injected simultaneously at a
fixed flow rate $Q_1$. Once the system has reached a steady state
($\text{ss}_1$), we abruptly change the flow rate to a different value
$Q_2$ and wait until a new steady state is established
($\text{ss}_2$). Finally, we change the flow rate back to its initial
value $Q_1$ and let the system evolve towards a third steady state
($\text{ss}_3$).  We have used different couples
$\left\{Q_1,Q_2\right\}$, listed in Table \ref{expparams}, to obtain
different magnitudes and signs of $\Delta Q=Q_1-Q_2$. We compare
$\text{ss}_1$ and $\text{ss}_3$ using three criteria: the average
pressure drop across the system, the fluid saturations, and the size distribution of the air
clusters.

\begin{table}[ht]
\centering 
\begin{tabular}{c c c || c c} 
\hline\hline 
Experiment & $Q_1 $ (mL/h) & $Q_2 $ (mL/h) & $\chi_{inlet}$ &
$\chi_{middle}$ \\ [0.5ex] 
\hline 
Exp.1  & 6.15 & 61.1 &  0.91 & 0.31 \\  
Exp.2  & 61.1 & 6.15 &  0.96 & 0.93 \\  
Exp.3  & 6.15 & 156 &  1.89 & 1.12 \\
Exp.4  & 156 & 6.15 &  2.86 & 2.31 \\
Exp.5  & 15.0 & 30.5 &  0.67 & 0.04 \\
Exp.6  & 6.15 & 61.1 &  0.19 & 0.04 \\
[1ex] 
\hline 
\end{tabular} 
\caption{\label{expparams} Experimental parameters and corresponding
  measurements. $Q_1$ and $Q_2$ refer to the total imposed flow rates (\textit{i.e.,} for 15 syringes). $\chi_{inlet}$ and $\chi_{outlet}$ are
  computed from the measured pressure drops according to
  Eq. \ref{eq:chi}. Note that experiments Exp.1 and Exp.6 are
  performed using the same parameters, illustrating the
  reproducibility of the results.}
\end{table}

From the pressures measured at the inlet, middle, and outlet of the
system, we compute the pressure drops $\Delta
P_i(t)=P_{inlet}(t)-P_{outlet}(t)$ and $\Delta
P_m(t)=P_{middle}(t)-P_{outlet}(t)$ (see Figure
\ref{setuptop}). Figure \ref{pressresults}(c) shows a plot of these
quantities for a typical experiment. After a transient characterized
by a linear increase, both pressure drops stabilize and fluctuate
around constant average values. This behavior is identical to what has
been observed previously \cite{tallakstad09} and defines steady state
1 ($\text{ss}_1$). The fluctuations of $\Delta P$ in steady state
reflect the dynamics of the non-wetting clusters -- transport,
mergings and snap-offs -- as the system explores different
configurations \cite{tallakstad09,tkrlmtf09}. Immediately after we
change the flow rate to the value $Q_2>Q_1$, both pressure drops
display a very rapid increase, and a new steady-state ($\text{ss}_2$),
characterized by higher values of $\Delta P$, is quickly
established. When we change the flow rate back to $Q_1$, we observe
again a rapid variation of the pressure drops towards a third plateau
defining steady-state 3 ($\text{ss}_3$).

Because experiments are performed at slow flow rates, and to obtain
enough statistics on the measurements, they must typically run for
several days. Indeed, the quality of the statistics is determined by
the ratio $N_{pv}\left(t\right)$ of the total volume of fluids
injected into the system to the total pore volume of the porous
matrix, namely:

\begin{equation}
 N_{pv}\left(t\right)=\frac{Q\left(t\right).t}{WLa\phi},
 \label{eq:Npv}
\end{equation}
where $Q$ denotes the total flow rate, $t$ denotes time and $W$, $L$, $a$
and $\phi$ are the width, length, thickness and porosity of the porous
matrix, respectively. Figure \ref{pressresults}(b) shows $N_{pv}$,
calculated from the imposed flow rates as a function of time, for a
typical experiment. In all the experiments presented here, the
duration of the steady states correspond to the injection of 0.3 to
1.4 pore volumes in the system. As a consequence, it is necessary to
refill the syringes one or several times in the course of an
experiment. This is systematically performed using the following
protocol: the outlet and inlets of the model are closed and the
syringe pump is stopped immediately. The refilling process takes
approximately 20 minutes. Because the vents used to close the model
are not perfectly air-proof, the pressure in the system relaxes
towards atmospheric pressure during this process, explaining the
sudden drop of $\Delta P$ observed on Figure
\ref{pressresults}(c). However, by looking at the pictures recorded
throughout the process, we have checked that this does not affect the
flow structure, which remains immobile over the duration of the
refilling procedure. Furthermore, we observe that once the syringe
pump is restarted, $\Delta P$ retrieves its initial value after a
delay originating from the compressibility of air. Thus this refilling
procedure does not affect the results of the experiments. In the
following, when necessary, we distinguish steady states obtained
before and after refiling using the subscripts a and b, respectively
(see Figure \ref{pressresults}(c)).

\subsection{History-independence of pressure drops}

\begin{figure}[htp]
\centering
\includegraphics[width=0.5\textwidth]{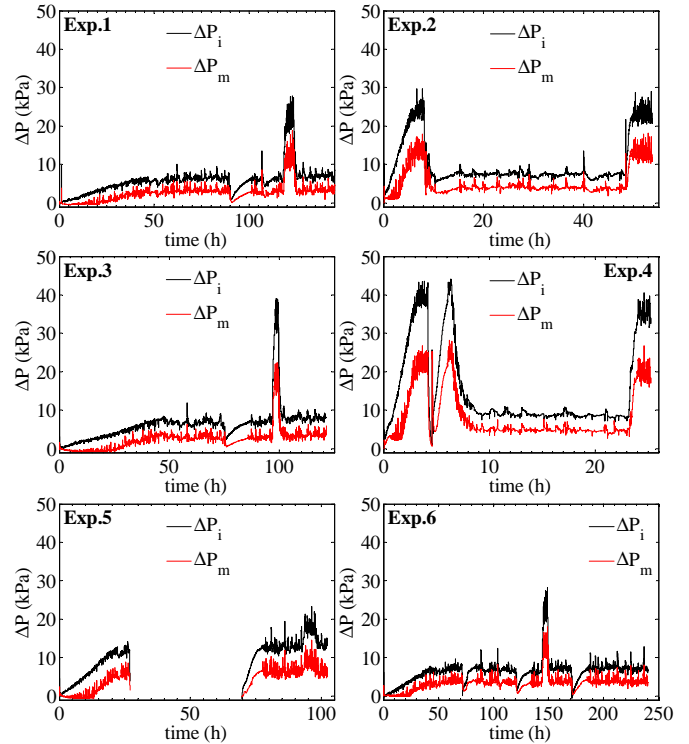}
\caption{Measured pressure drops for the 6 experiments listed in Table
  \ref{expparams}.}\label{pressureDrops}
\end{figure}

Figure \ref{pressureDrops} shows the temporal evolution of the
pressure drops $\Delta P_{i}$ and $\Delta P_{m}$ for the 6 experiments
listed in Table \ref{expparams}. In all cases, we observe a behavior
similar to what has been described in the previous section, namely
rapid variations of $\Delta P$ upon changes of flow rate between
plateaus characterizing steady states. Most importantly, we observe
that whatever the value and sign of $\Delta Q$, the pressure drops are
similar in $\text{ss}_1$ and $\text{ss}_3$. This suggests that the steady-state pressure drop $\Delta
P_{\rm ss}$ only depends on the imposed flow rate, and not on the history
of the system. Note also that for a given flow rate, $\Delta P_{\rm ss}$
values are reproducible from one experiment to the next, regardless of
how steady state has been reached.

To obtain a quantitative indication of this history-independence, we
compute average steady-state pressure drop values $\langle \Delta
P_{\rm ss} \rangle$ over the duration of each steady-state, thus
over periods of time corresponding to the injection of 0.3 to 1.4 pore
volumes in the system, as already indicated earlier. For each
experiment, we compare the variation of $\langle \Delta P_{\rm ss} \rangle$
between $\text{ss}_1$ and $\text{ss}_3$ to the fluctuations of $\Delta
P_{\rm ss}$ within these two states. For this we compute the ratio:

\begin{equation}
 \chi=\left|\frac{ \langle \Delta P_{ss_1} \rangle - \langle \Delta
   P_{ss3} \rangle}{0.5\left(\sigma\left(\Delta
   P_{ss_1}\right)+\sigma\left(\Delta P_{ss_3}\right)\right)}\right|,
 \label{eq:chi}
\end{equation}
where $\langle \cdots \rangle$ represents a temporal average over the
duration of a steady-state and $\sigma(\Delta P_{\text{ss}_1}) \approx
\sigma(\Delta P_{\text{ss}_3})$ are the standard deviations of $\Delta
P_{\rm ss}$ in $\text{ss}_1$ and $\text{ss}_3$. The values of $\chi$ computed
from the inlet and middle pressure drops for the different experiments
are reported in Table \ref{expparams}. As can be observed, almost all
of these values are smaller than one. Slightly higher values are
observed in Exp.3 and Exp.4. However, our experimental temperature data suggests that temperature-induced viscosity fluctuations most likely explain this fact.
Indeed, the largest temperature variations both between $\text{ss}_1$ and $\text{ss}_3$ and within $\text{ss}_1$ or $\text{ss}_3$ occur for these two experiments.
Therefore, we find that within the precision of our measurements, the steady-state pressure drop values
are history-independent.

\subsection{History-independence of saturation and non-wetting 
cluster size distributions}

Pressure measurements suggest that the observed steady-state only
depends on the imposed flow rate. However, these measurements do not
give us detailed spatial information. Thus, we now analyze the images
of steady-state flow patterns. As explained earlier (see Figure
\ref{flowimage}), the flow structure consists of clusters of the
non-wetting phase, air, surrounded by the wetting viscous
water-glycerol solution. As air clusters are transported through the
porous medium, they are fragmented or merged, giving many different
realizations of the flow pattern. However, within steady-state, the saturation
and the size distribution of air clusters remain constant on average
\cite{tallakstad09}. Thus, similarly to what we have done for pressure
drops, we compute average steady-state saturations and cluster size
distributions and compare their properties in $\text{ss}_1$ and
$\text{ss}_3$. To obtain a good statistics on these quantities, we
have chosen the frame rates so that successive images sample different
realizations of the flow pattern. All averages are computed over
series of images, typically 100, spanning a time range corresponding
to the injection of $1/5$ of the pore volume at the minimum.

\begin{figure}
\centering
\includegraphics[width=0.5\textwidth]{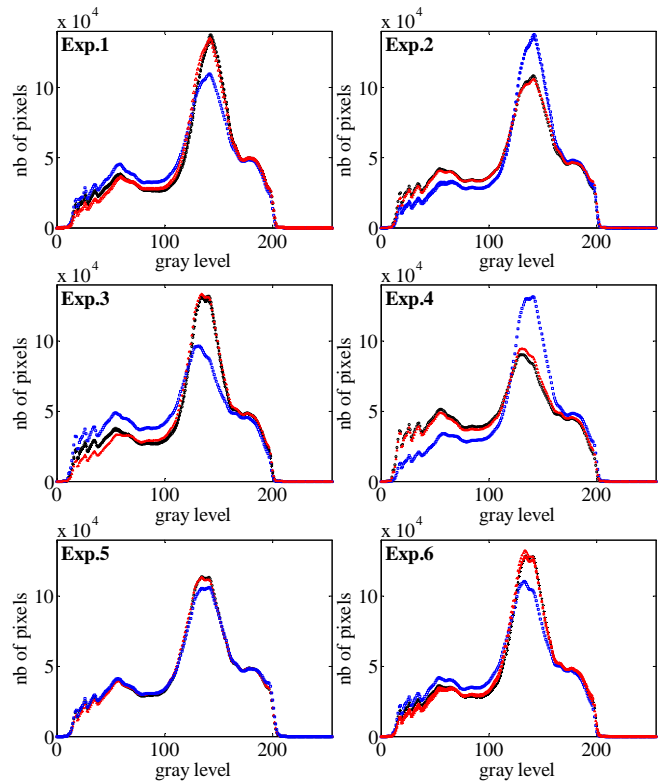}
\caption{Gray-scale image histograms averaged over steady-state images
  for the 6 experiments listed in Table \ref{expparams}. Note that no
  image processing has been applied before obtaining these
  histograms. Data for steady-states $ss_1$, $ss_2$ and $ss_3$ are
  represented in black, blue and red, respectively. Different symbols
  refer to averages performed over different series of 100 images,
  namely: $\circ$, $\cdot$ for images in $ss_{1a}$, $+$, $\times$ in
  $ss_{1b}$, $\square$, $\lozenge$ in $ss_{2}$ $\vartriangle$,
  $\triangledown$ in $ss_{3a}$ and $\triangleleft$, $\triangleright$
  in $ss_{3b}$. }\label{rawHistos}
\end{figure}

As explained in Section \ref{section-setup}, the gray-scale histograms
of the raw images give us a direct indication of the proportion of the
two phases in the system, thus of saturation (see Figure
\ref{flowimage}). Figure \ref{rawHistos} shows average steady state
histograms for the 6 experiments listed in Table \ref{expparams}. It
is clear on all these graphs that the histograms corresponding to
$\text{ss}_1$ and $\text{ss}_3$ can be distinguished from those
corresponding to $\text{ss}_2$: indeed, as expected, the saturation
depends on the flow rate, as reflected by different air and liquid
``peak'' heights on the histograms. However, histograms are similar in
$\text{ss}_1$ and $\text{ss}_3$, suggesting that the steady-state saturation is
also history-independent. It is important to note that histograms are
directly obtained from raw images without prior processing, which
excludes eventual artifacts due to image processing. However, they do
not give us any information about the spatial repartition of the two
phases in the system.

To obtain this information, we process the images to identify air
clusters and compute their size distributions. Image processing is
performed using ImageJ~\cite{imageJ}. Raw images
are thresholded to obtain binary (black and white) images on which we
run a standard particle analysis algorithm\footnote {We use the
  ``Particles4'' ImageJ plugin written by G. Landini (see
  http://www.dentistry.bham.ac.uk/landinig)} to identify air clusters
and measure their sizes $n$.  From steady-state images, we compute the
normalized probability density functions of $n$, \textit{i.e,} non-wetting
cluster size distributions $\langle p(n) \rangle$, where $\langle
\cdots \rangle$ represents an average over a series of $\approx 100$
images.

\begin{figure}
\centering
\includegraphics[width=0.5\textwidth]{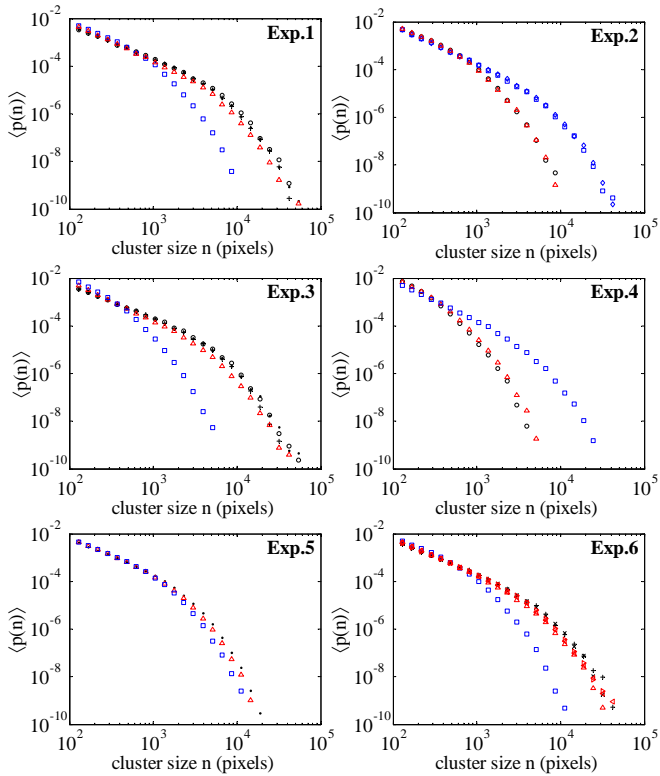}
\caption{Average cluster size distributions $\langle p(n) \rangle$
  computed from steady-state images. We use the same symbols and
  colors as on Figure \ref{rawHistos}}\label{distribs}
\end{figure}

Figure \ref{distribs} shows the distributions $\langle p(n) \rangle$
computed for the 6 experiments listed in Table \ref{expparams}. These
distributions typically display a power-law-like behavior with a
cutoff at large cluster sizes \cite{tkrlmtf09,tallakstad09}. As
mentioned by Tallakstad \textit{et al.} \cite{tallakstad09}, the
obtained distribution is affected by threshold values, which must thus
be carefully chosen using visual inspection. Here, we focus on the
variations of the distribution with the history of the
system. Therefore, the most important requirement is that the image
processing procedure is used consistently throughout one
experiment. To avoid possible bias due to variations of illumination
in the room, the experimental setup is isolated behind a dark
curtain. The camera exposure time and aperture are the same for all
experiments, and we use the same thresholding parameters, carefully
chosen by visual inspection, for all experiments. This allows us to
compare images obtained in $\text{ss}_1$, $\text{ss}_2$ and
$\text{ss}_3$ for a given experiment and from one experiment to the
other in a meaningful way. Similarly to what we observed for the
histograms, it is possible to distinguish the $\text{ss}_1$ and
$\text{ss}_3$ distributions from those corresponding to
$\text{ss}_2$ (see Figure \ref{distribs}). This is coherent with the
results of previous studies indicating that distributions are shifting
towards higher clusters sizes when increasing the flow rate
\cite{tkrlmtf09,tallakstad09}. However, the $\text{ss}_1$ and ${ss}_3$
distributions are similar, meaning that the steady-state non-wetting cluster size
distributions are history-independent. We have checked that whereas
varying the threshold values affects the distributions, typically by
shifting then towards lower or higher cluster sizes, it does not
modify the results in terms of history independence.

The experimental boundary conditions
imposed that the controlled flow variables were the total flow rate
and the fractional flows. In the next Section, we turn to numerical
simulations to further investigate the history-dependence of the
steady-state for different boundary
conditions, as well as higher $\rm Ca$ values and different viscosity ratios $\rm M$ .

\section{The network model}\label{section-network-model}
\begin{figure}
\centerline{\hfill
\includegraphics[scale=0.28,clip]{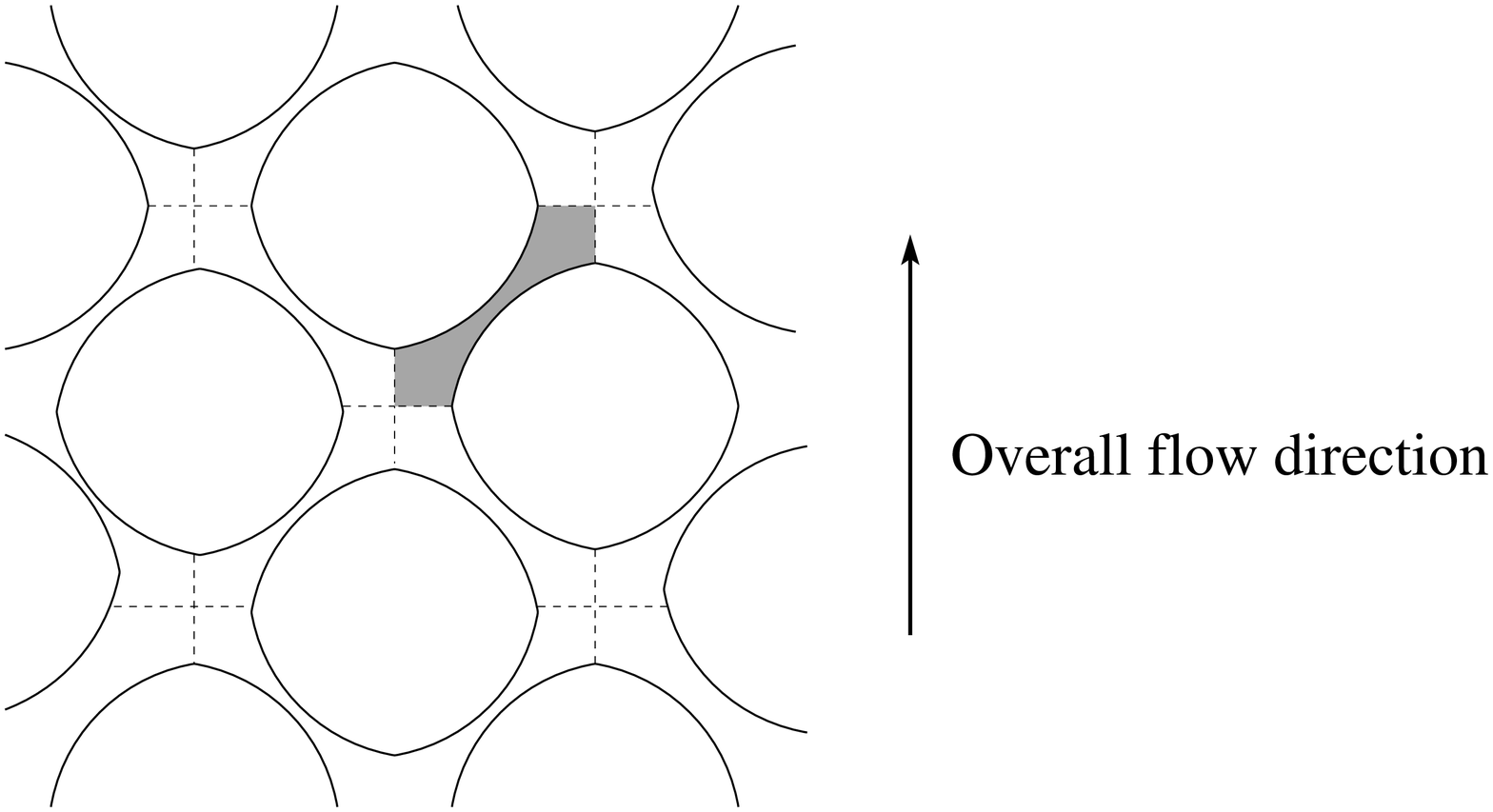}\hfill
}
\caption{\label{demo} Illustration of the network formed by tubes that
  are connected to each other through nodes where the dashed lines are
  intersect. One single tube is colored gray. Spherical glass beads
  makes the tubes as hour-glass shaped.}
\end{figure}

The two-dimensional experimental porous medium is
modeled by a network of tubes orientated at $45^\circ$ with respect to
the overall flow direction. The tubes (or links) intersect at the
vertices (or nodes) of the network with coordination number 4. The
nodes are considered to have no volume, so the tubes consist of both the
pore and throat volumes. The network is illustrated in Figure
\ref{demo}. The disorder due to the random position of the glass beads
in the experiment is introduced in the model by choosing the radius
$r$ of each tube randomly from a uniform distribution of random
numbers in the range $[0.1l, 0.4l]$, where $l$ is the length of a
tube. In order to incorporate the shape of the pores in between
spherical glass beads, each tube is considered as hour-glass shaped
which introduces the capillary effect in the system. The network
transports two immiscible fluids, one of which is more
wetting than the other with respect to the pore walls. The fluids are
separated by menisci and we obtain the capillary pressure $p_c$ at a
meniscus inside the hour-glass shaped tubes from a modified form of
Young-Laplace law \cite{amhb98, d92},
\begin{equation}
\displaystyle
p_c = \frac{2\gamma \cos \theta}{r}\big[1 - \cos\frac{2\pi x}{l}\big]\;,
\label{pc}
\end{equation}
where $x$ is the position of the meniscus and $\gamma$ is the
interfacial tension between the fluids. $\theta$ is the wetting angle
and we consider perfectly wetting conditions, \textit{i.e.} $\theta$
is either $0^\circ$ or $180^\circ$. The flow is driven by maintaining
a constant total flow rate $Q$ throughout the network which introduces
a global pressure drop. The local flow rate $q$ in a tube with a
pressure difference $\Delta p$ between its two ends follows the Washburn equation of capillary flow \cite{w21,amhb98}
\begin{equation}
\displaystyle q = -\frac{a k}{\mu_{\rm eff}(s_{\rm nw})
  l}\left(\Delta p - \sum p_c\right)\;,
\label{wb}
\end{equation}
where $k=r^2/8$ is the permeability for cylindrical tubes. Any other cross-sectional shape of the tubes will lead only to an overall geometrical factor. Here
$a$ is the cross-sectional area of the tube and $\mu_{\rm eff}(s_{\rm
  nw})$ is the volume average of the viscosities of the two phases
present inside the tube. Hence, it is a function of the local non-wetting
saturation $s_{\rm nw}$ in that tube. The sum over $p_c$ runs over all
the menisci inside the tube. The property that the fluid flux through
every node will be zero is used to obtain the local pressures at the
nodes. The set consisting of one equation (\ref{wb}) per tube,
together with the Kirchhoff equations balancing the in and out flow at
each node are then solved using the Cholesky factorization or the
conjugate gradient method. The system is then integrated in time using
an explicit Euler scheme. Inside a tube all menisci move with a speed
determined by $q$. When a meniscus reaches the end of a tube, new
menisci are formed in the neighboring tubes. In each link, a total
maximum number of menisci is allowed to form. When the number is
exceeded, two nearest menisci are merged keeping the volume of each
fluid conserved. Here, we have considered a maximum of
$4$ menisci in one tube (\textit{i.e.}, $2$ non-wetting bubbles), as
it is not very likely to form a lot of menisci in one pore as seen
from the experimental observations. Further details of the model and
how the menisci are moved can be found in \cite{amhb98, kah02}.

In order to reach the steady state in the simulation, we considered
two different approaches. The conventional way is to implement the
bi-periodic (BP) boundary condition, where the links at the outlet row
are connected to the inlet links, so that the network acquires a
toroidal topology \cite{kh02}. In this case the simulation can go
forever, regardless of whether one fluid percolates or not. However,
in order to keep the flow going, the global pressure gradient is
maintained by considering an invisible cut through the system in terms
of the pressure. Since the system is closed with this boundary condition,
the individual fluid volumes remain constant throughout the
simulation. The non-wetting saturation $S_{\rm nw}=V_{\rm nw}/V$ is therefore an
independent parameter here, along with the total flow rate $Q$,
whereas the non wetting fractional flow $F_{\rm nw}=Q_{\rm nw}/Q$
fluctuates over time.

Implementing the bi-periodic (BP) boundary condition is of course
impossible in the experiments. As described before, in the experimental
setup, two fluids are injected at one edge of the Hele-Shaw cell
through a series of alternate inlets and the opposite edge is kept
open. In this case, the flow rates of the two fluids can be controlled independently.
Thus, the control parameters are the total
flow rate $Q$ and the fractional flow $F_{\rm nw}$, whereas the
saturation $S_{\rm nw}$ fluctuates. In order to have a close emulation
of the experimental ensemble, we also implement the open boundary
conditions (OB) in the simulations, where the individual flow
rates at the inlet links are controlled. Therefore, in OB, the system is open
in the direction of total flow while we consider periodic
boundary conditions in the direction perpendicular to the overall flow.

\section{Simulation Results}

\begin{figure}
\centerline{\hfill
\includegraphics[scale=0.22]{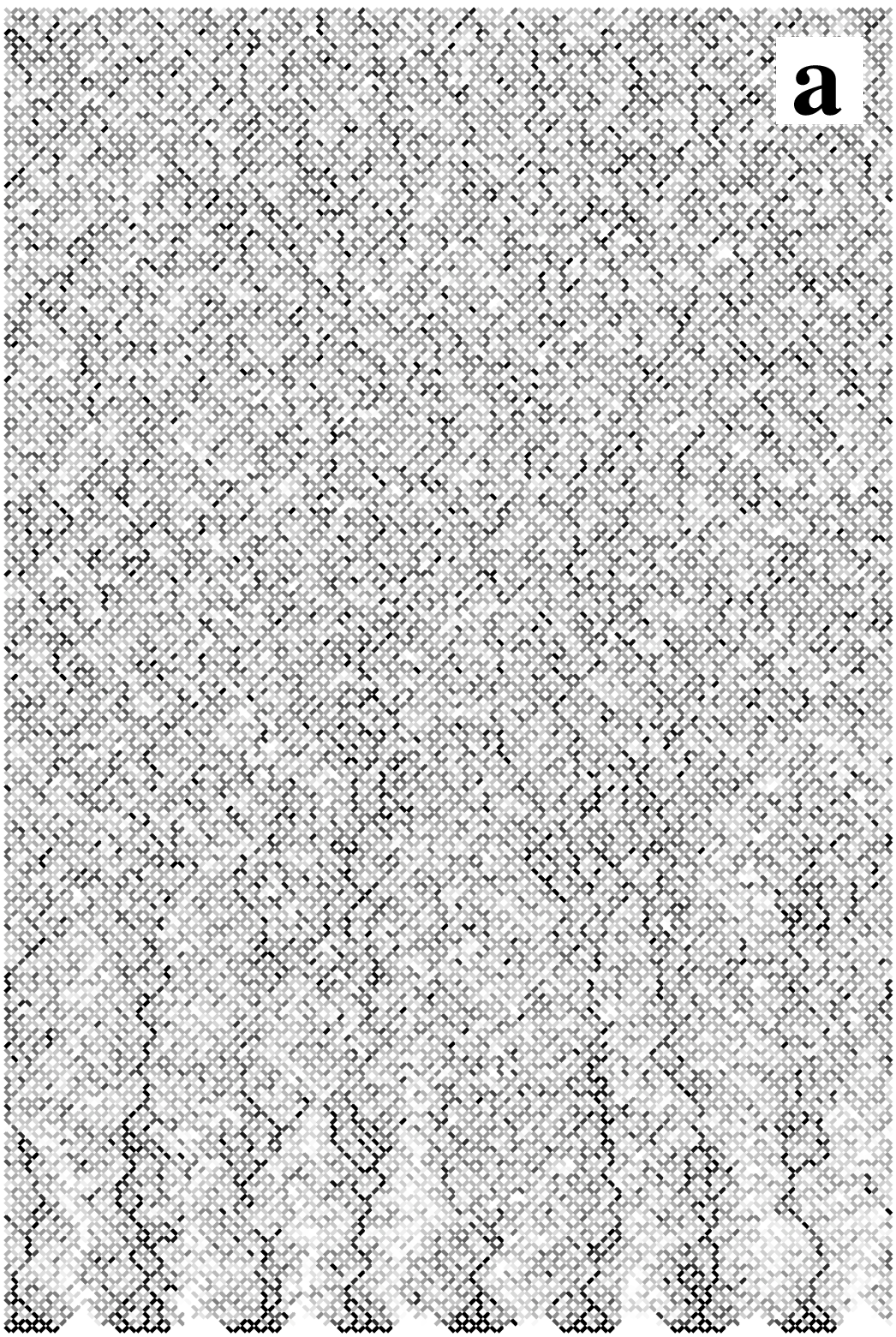}\hfill
\includegraphics[scale=0.22]{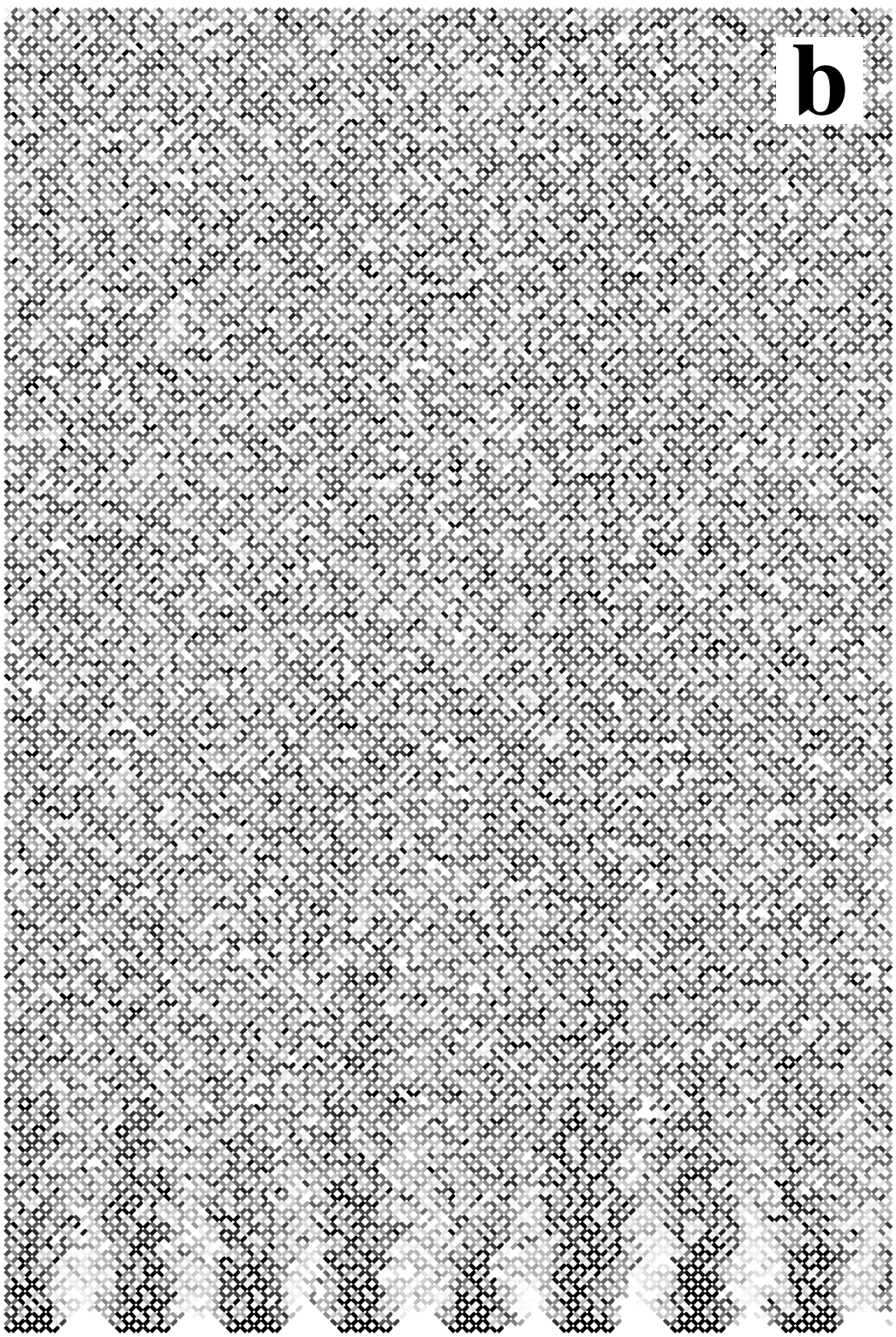}\hfill
\includegraphics[scale=0.22]{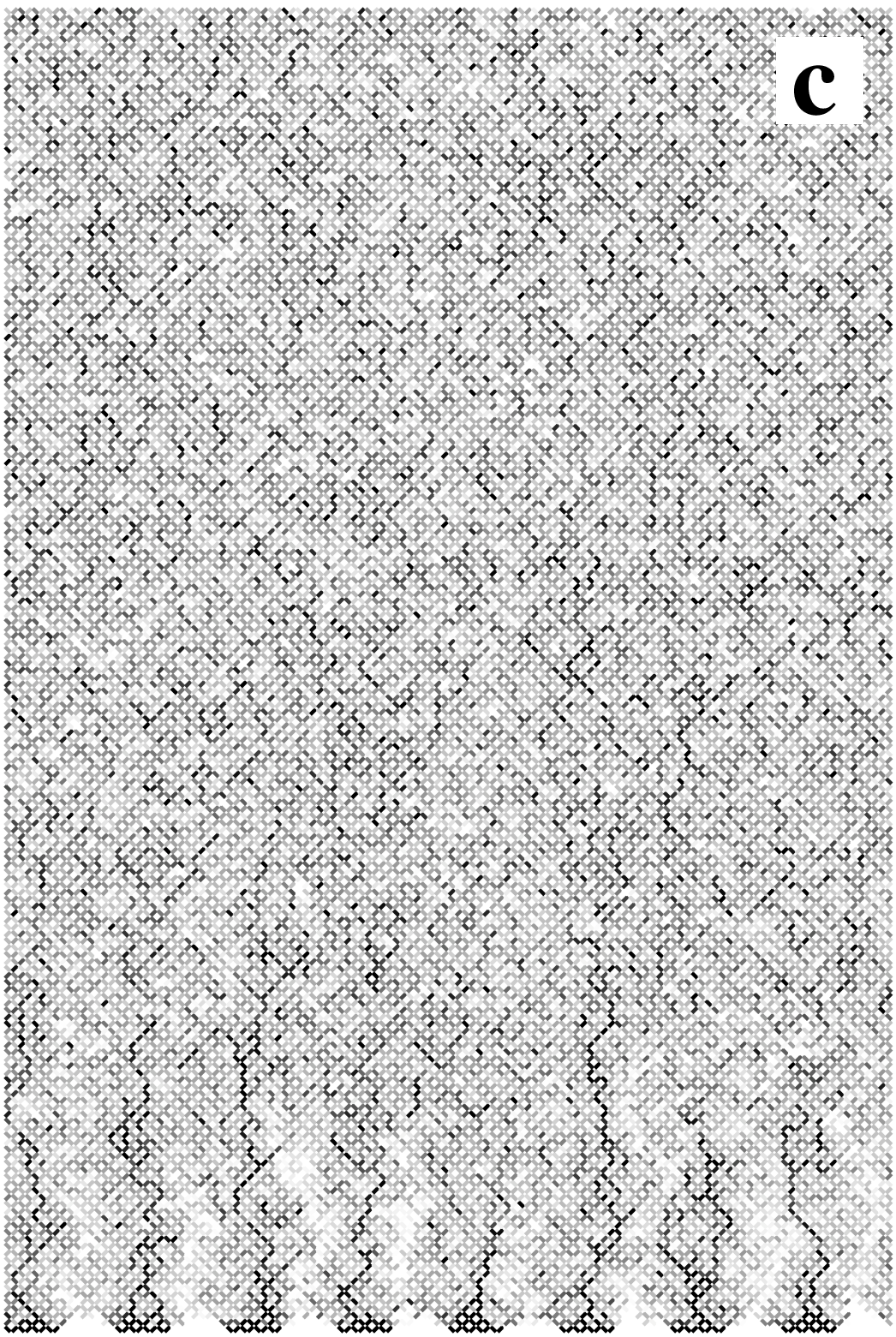}\hfill
}
\medskip
\centerline{\hfill
  \includegraphics[scale=0.22]{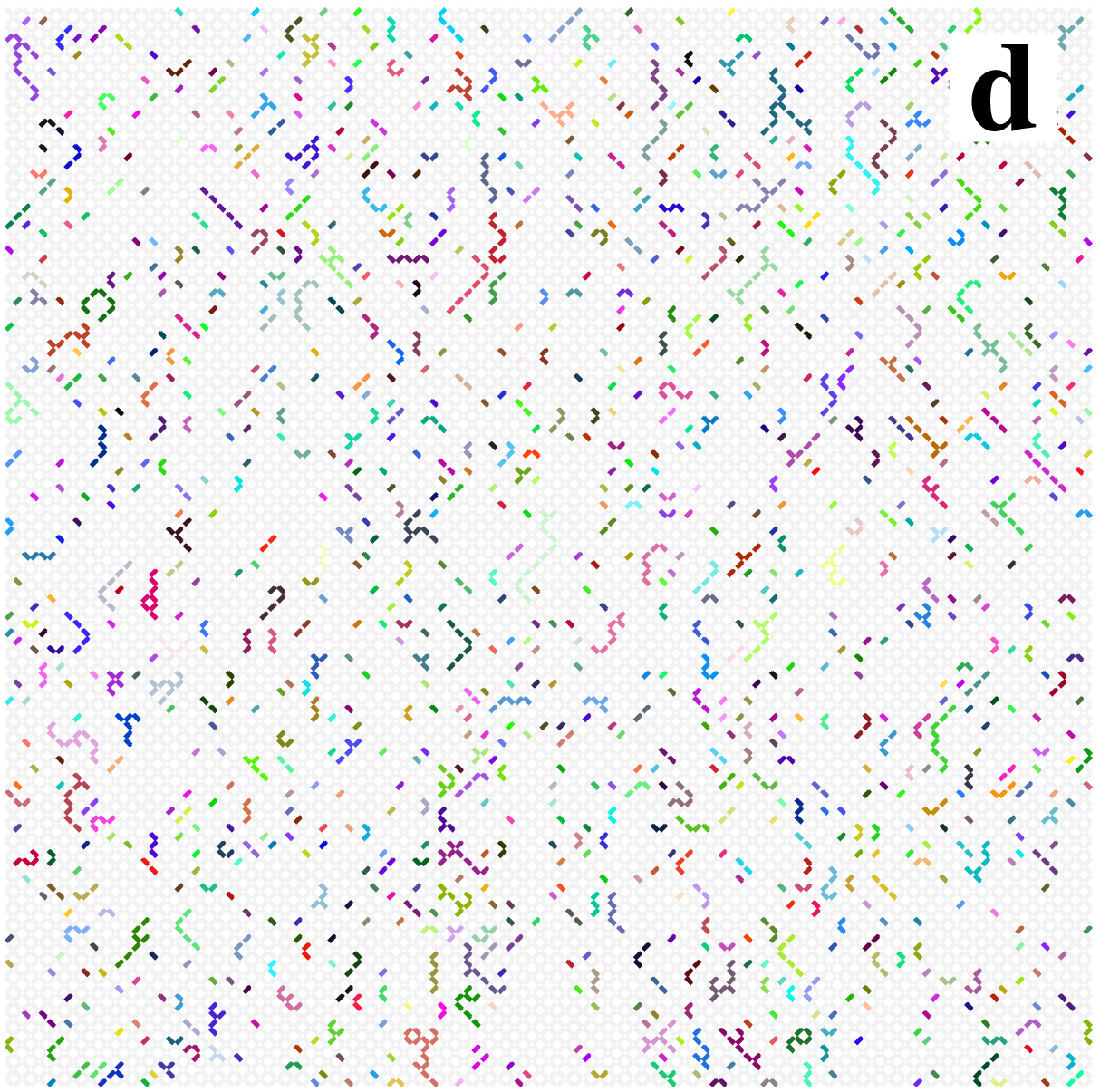}\hfill
  \includegraphics[scale=0.22]{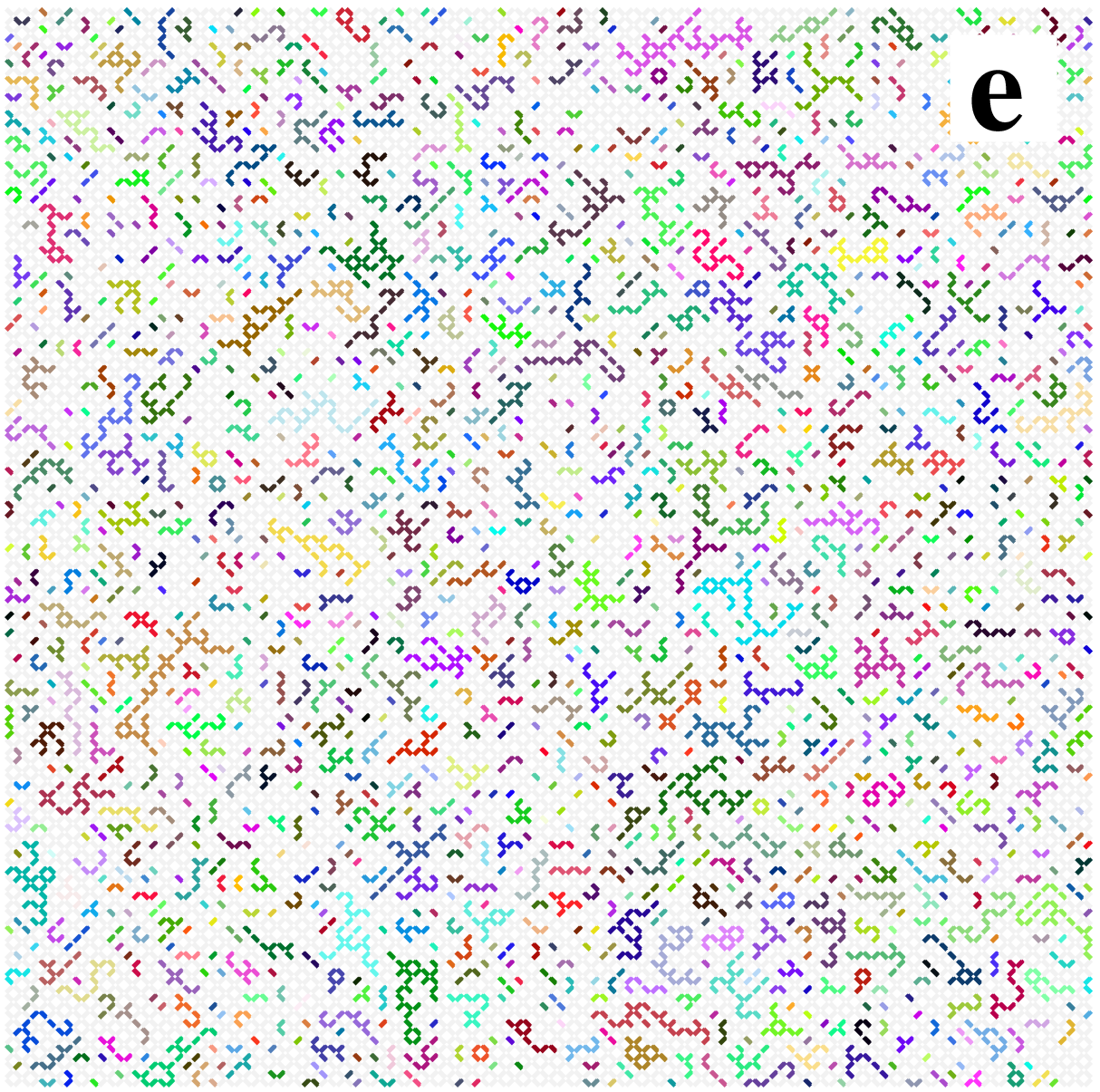}\hfill
  \includegraphics[scale=0.22]{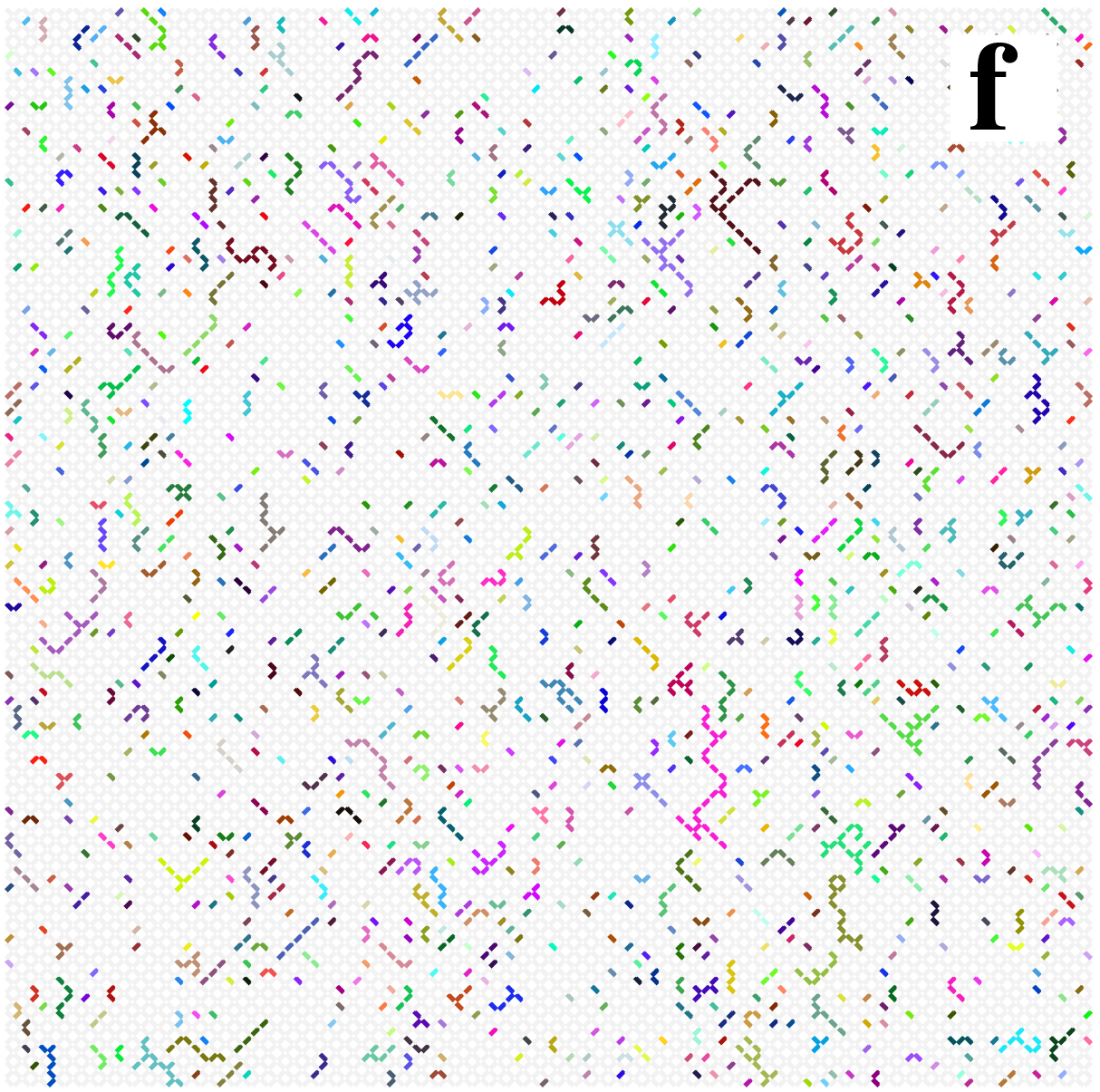}\hfill
}
\centerline{\hfill $\rm Ca = 9.15\times 10^{-3}$ \hfill $\rm Ca =
  2.88\times 10^{-2}$ \hfill $\rm Ca = 9.15\times 10^{-3}$ \hfill}
\caption{\label{clspic_ob} Steady-state fluid morphology over the
  network in OB for $\rm M=1$. The figures are drawn in the similar way as that of
  BP.}
\end{figure}

Simulations are performed considering networks of $256\times 256$
links for BP and $128\times 192$ links for OB. In order to avoid any
traces from the inlets in OB, only a $128\times 128$ segment of the
network towards the outlets is considered for the analysis (see Fig.~\ref{clspic_ob}). Each link
has a length of $1 \text{ mm}$, which is equal to the bead diameter used in the
experiments. Most of the simulations are performed for the viscosity ratio
$\rm M = 1$. Three different capillary numbers, $\rm Ca_1 = 1.92\times
10^{-5}$, $\rm Ca_2 = 9.15\times 10^{-3}$ and $\rm Ca_3 = 2.88\times
10^{-2}$ are considered for BP. For OB, the capillary numbers
considered are $\rm Ca_1 = 3.2\times 10^{-3}$, $\rm Ca_2 = 3.2\times
10^{-2}$ and $\rm Ca_3 = 7.0\times 10^{-2}$. For OB we choose a similar
fractional flow $F_{\rm nw} = 0.5$ as in the experiments. In BP, we
run the simulation for the saturation $S_{\rm nw} = 0.74$ which we
find close to the critical point for the range of parameters we
considered here \cite{rh06}. We will report only one set of
simulations for $\rm M = 10^{-4}$ with BP for a system with $128\times
128$ links and $S_{\rm nw} = 0.5$ with capillary numbers $5.06 \times
10^{-2}$ and $1.24 \times 10^{-1}$, as the conjugate gradient solver
converges very slowly for viscosity unmatched fluids, making the
simulation very computationally expensive.

To investigate the history
dependence, we use a procedure analogous to the experimental one (see Sec.~\ref{sec-exp-procedure}): the simulation is started from the initial condition with
a capillary number setting a constant flow rate $Q_1$. Once a
steady state $\rm ss_1$ is reached, the capillary number is altered
to a different value setting another constant flow rate $Q_2$, and
this new flow rate is maintained until a different steady state
$\rm ss_2$ is reached. Finally, the capillary number is set
back to the initial value and the system is
allowed to evolve towards a third steady state $\rm ss_3$.
Each simulation thus involves two different capillary
numbers among $\rm Ca_1$, $\rm Ca_2$, and $\rm Ca_3$, therefore $6$
different simulations have been performed for each boundary
condition. The steady states are compared with three different
criteria -- the average pressure drop, the distribution of fluid
saturation over the system, and the non-wetting cluster size
distribution. All the measurements are averaged over $50$ to $100$
configurations in the steady-state and $5$ to $10$ different
realizations of the network.

\begin{figure}
\centerline{\hfill
\includegraphics[scale=0.22]{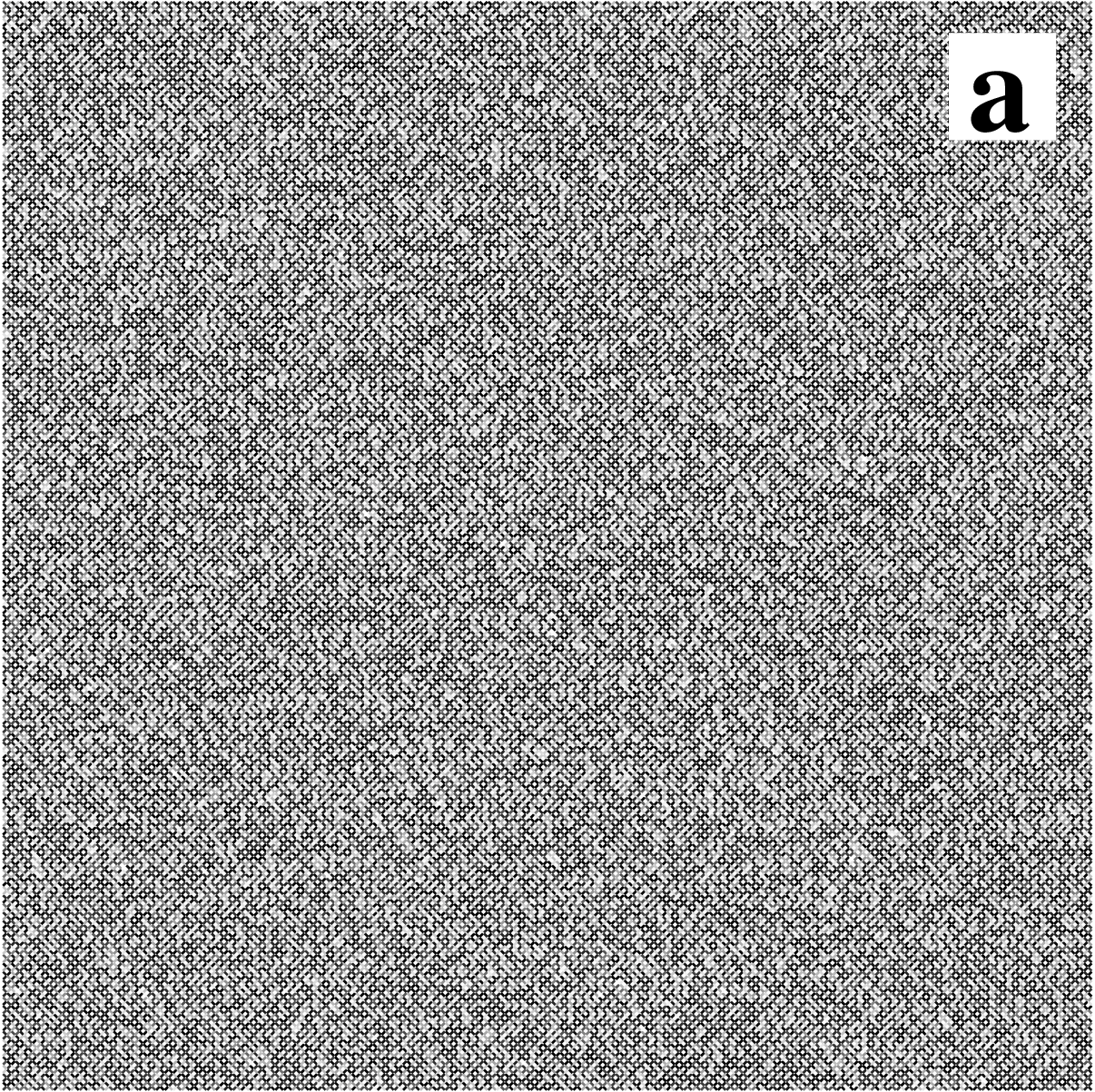}\hfill
\includegraphics[scale=0.22]{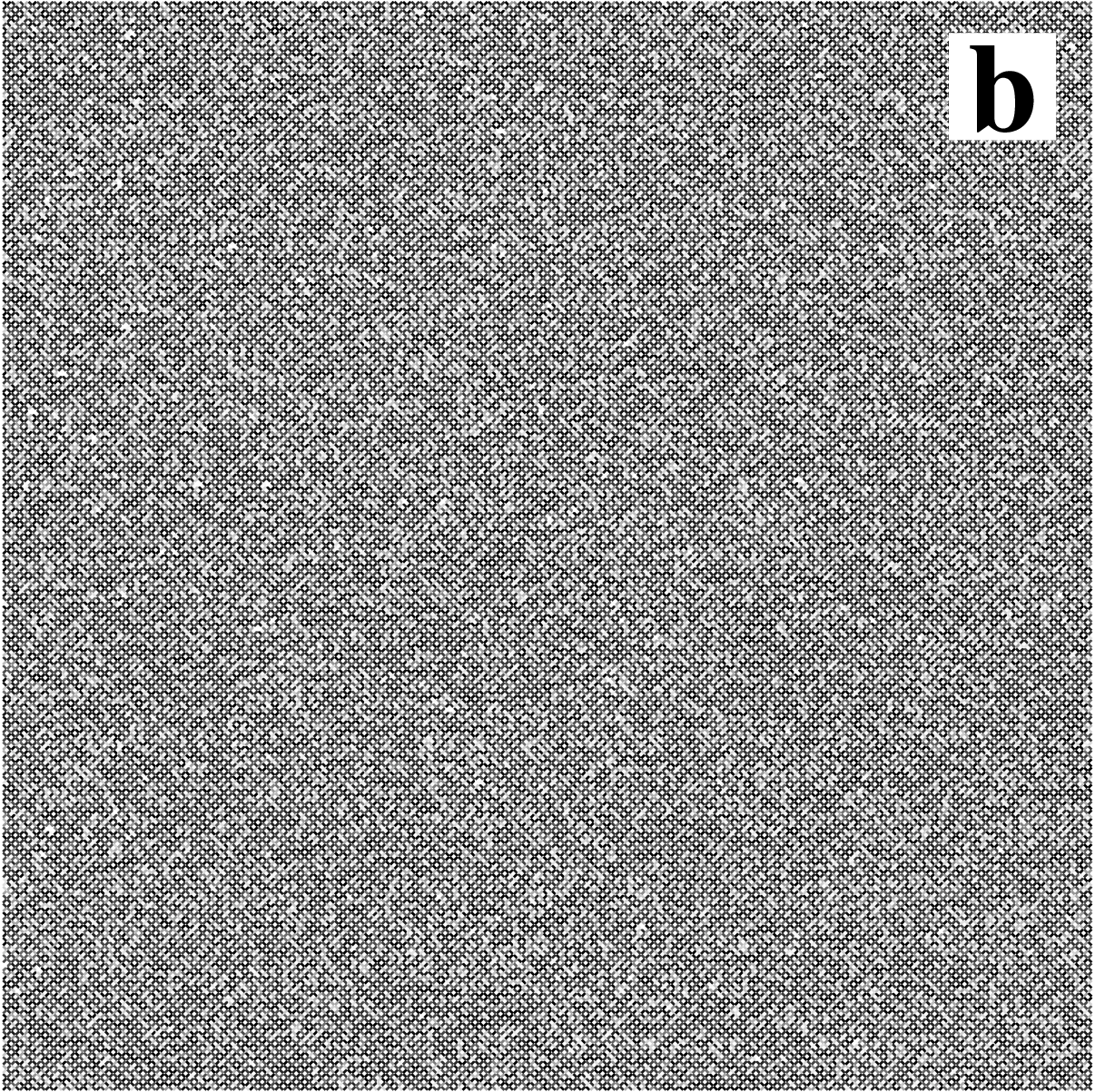}\hfill
\includegraphics[scale=0.22]{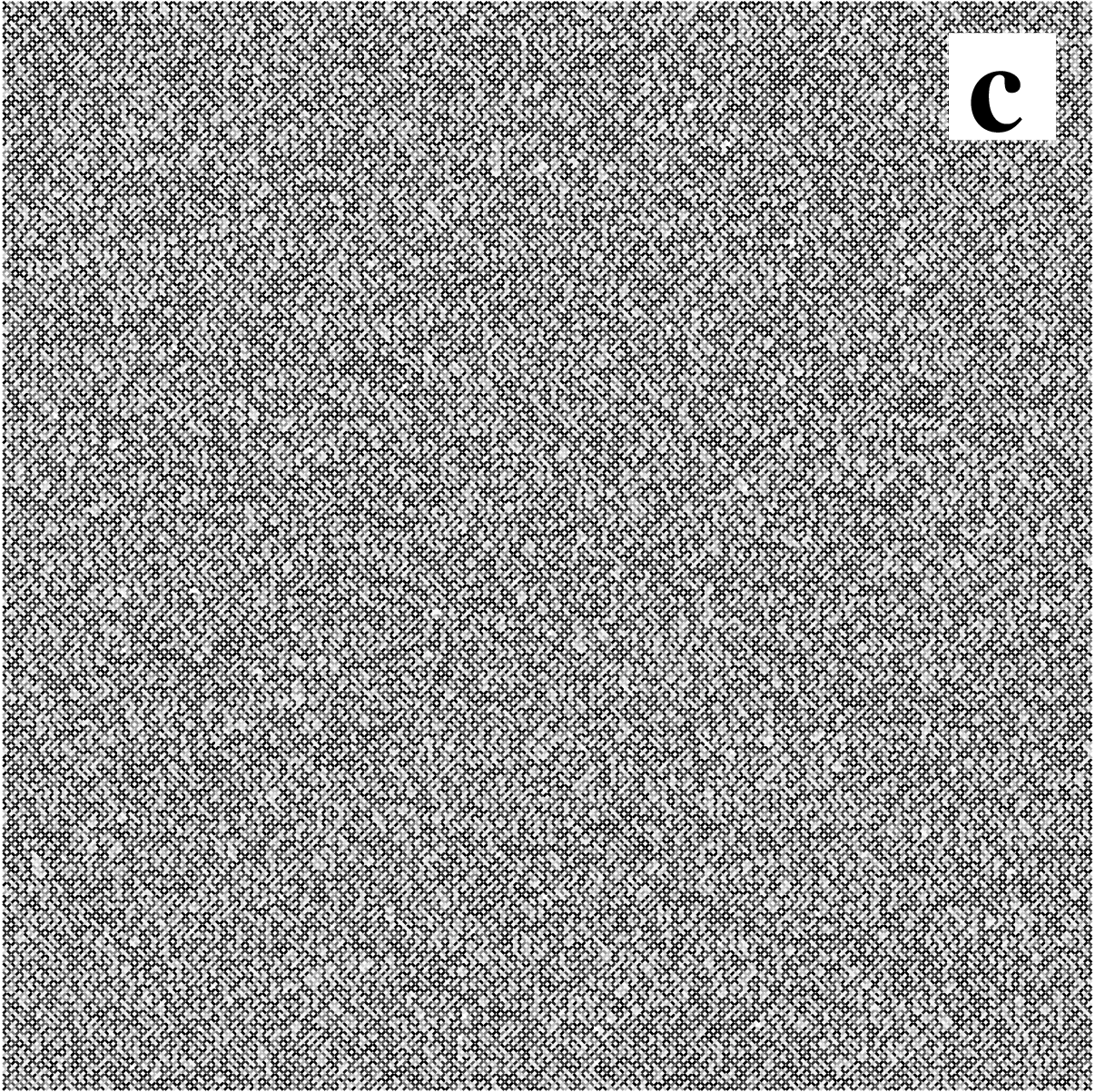}\hfill
}
\medskip
\centerline{\hfill
  \includegraphics[scale=0.22]{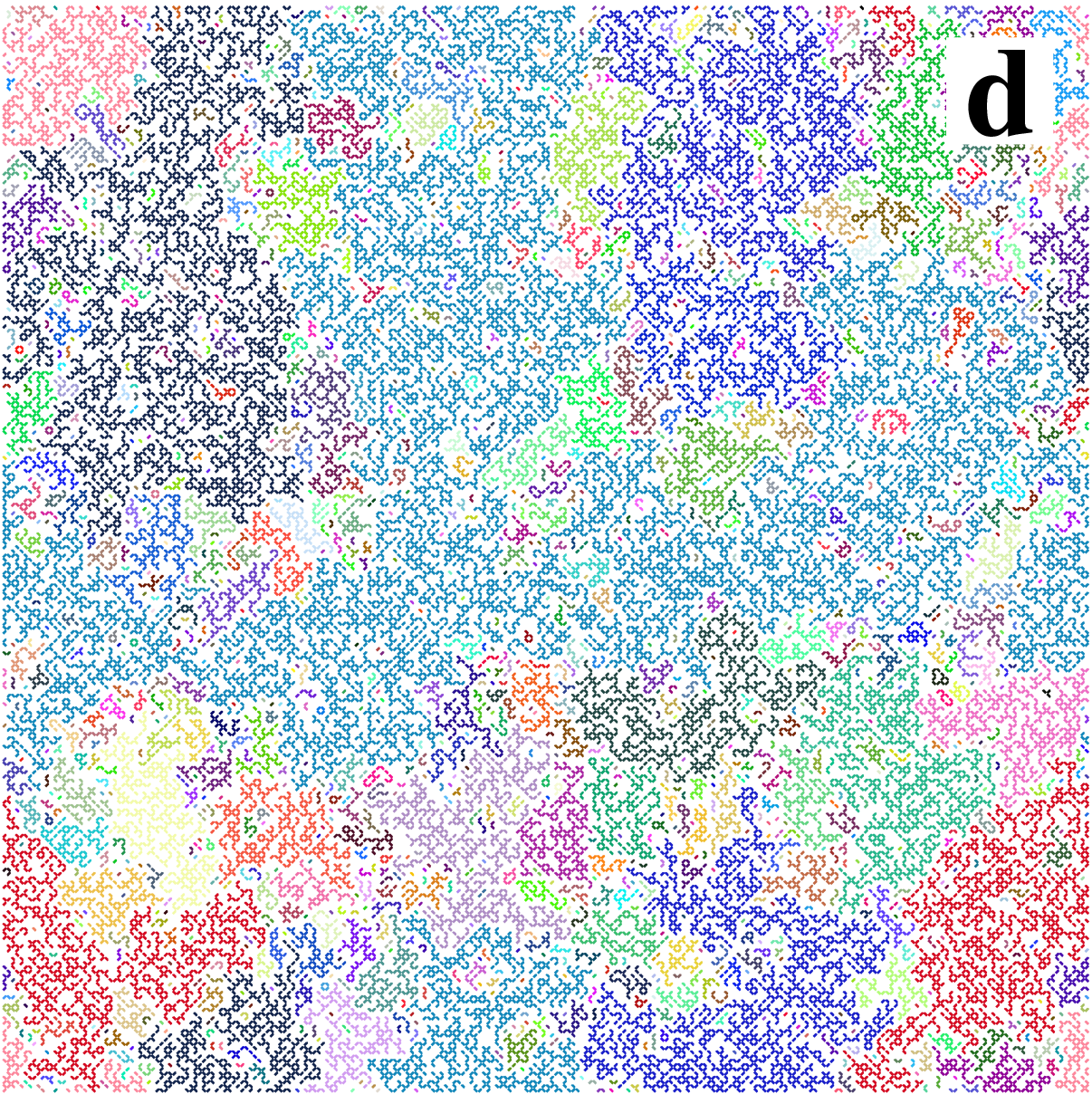}\hfill
  \includegraphics[scale=0.22]{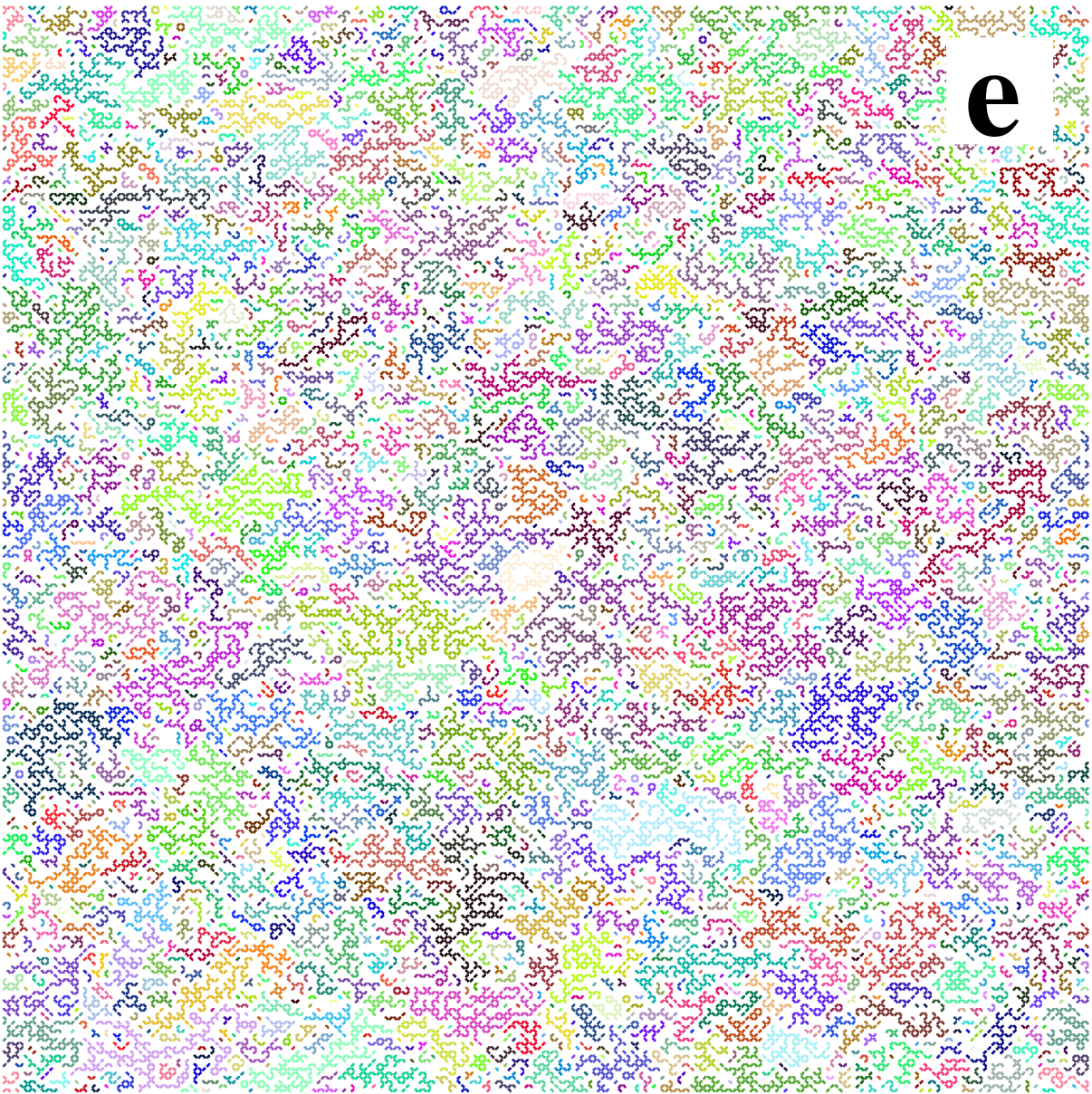}\hfill
  \includegraphics[scale=0.22]{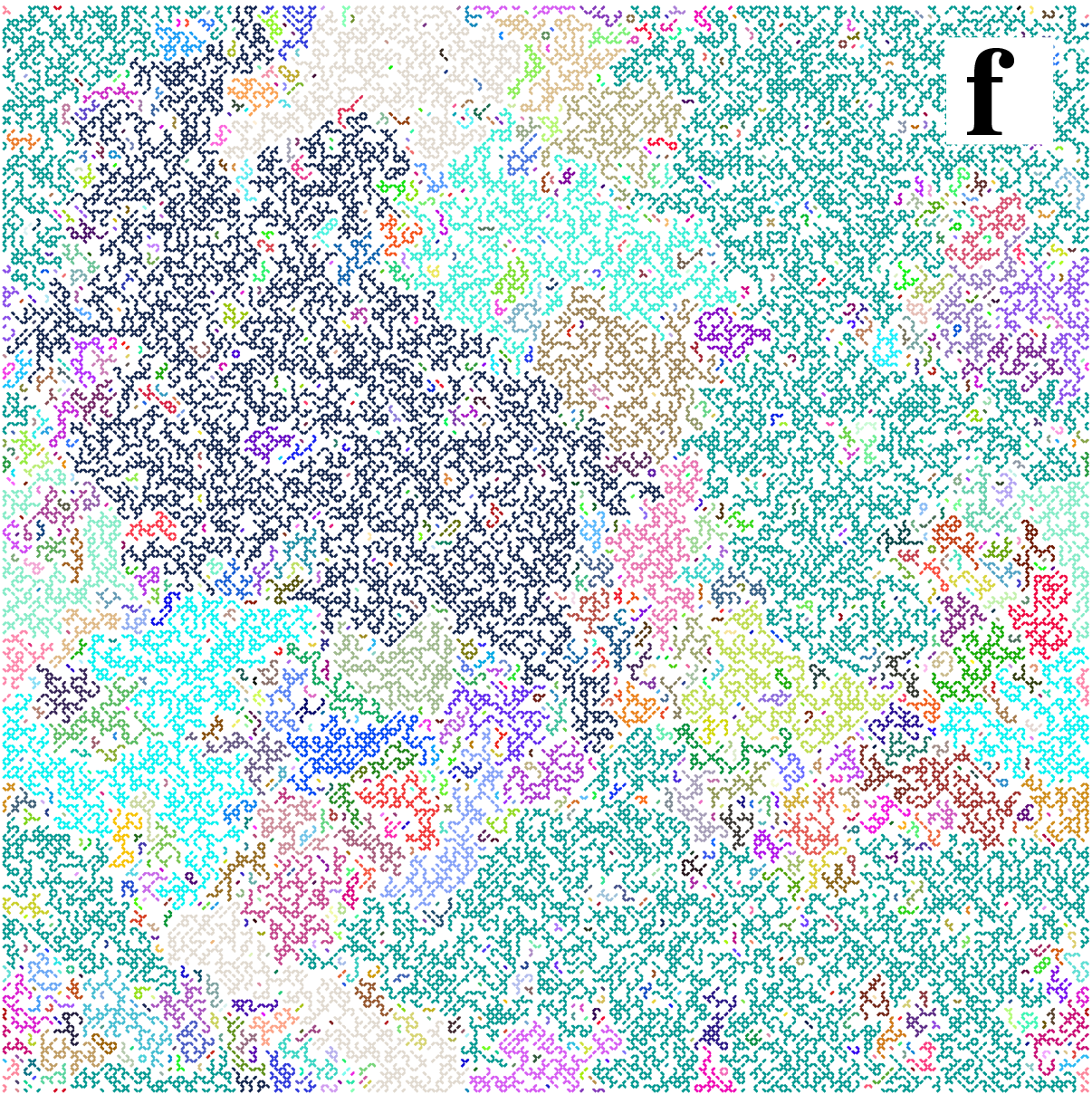}\hfill
} 
\centerline{\hfill $\rm Ca = 9.15\times 10^{-3}$ \hfill $\rm Ca =
  2.88\times 10^{-2}$ \hfill $\rm Ca = 9.15\times 10^{-3}$ \hfill}
\caption{\label{clspic_bp} Typical steady-state fluid morphology over
  the network in BP for $\rm M=1$ and $S_{\rm nw}=0.74$. The distribution of non-wetting fluid saturation
  inside the links in the steady-states $\rm ss_1$, $\rm ss_2$ and
  $\rm ss_3$ are illustrated in (a), (b) and (c) respectively and the
  corresponding capillary numbers are indicated under each column. The
  gray-scales from black to white correspond to the non-wetting
  saturation from $1$ to $0$ inside a link. The non-wetting clusters
  identified by the Hoshen-Kopelman algorithm are shown by different
  random colors in (d), (e) and (f).}
\end{figure}

Fluid morphologies for a typical simulation in the three
steady states $\rm ss_1$, $\rm ss_2$ and $\rm ss_3$ are shown on
Figures \ref{clspic_bp} and \ref{clspic_ob} for BP and OB
respectively. Here the simulation starts from $\rm Ca = 9.15\times
10^{-3}$ to reach $\rm ss_1$, then it is altered to $\rm Ca =
2.88\times 10^{-2}$ to reach $\rm ss_2$, and then it is again turned
back to the initial $\rm Ca = 9.15\times 10^{-3}$ to reach $\rm
ss_3$. Figures \ref{clspic_bp} and \ref{clspic_ob} show the distribution of saturation over the network in these three
steady-states in (a), (b) and (c) respectively, where the
gray-scales from black to white correspond to $s_{\rm nw} = 1$ to $0$
inside a link. For BP, it is not possible to see any difference in the gray-scale saturation distributions, as the system is closed and the total saturation is conserved. However, as we will see, identifying the clusters allows us to distinguish $\rm ss_1$ and $\rm ss_3$ from $\rm
ss_2$ (see Figs. \ref{clspic_bp} (d), (e) and (f)). In OB, the saturation distributions look similar in $\rm ss_1$
and $\rm ss_3$ whereas $\rm ss_2$ shows more non-wetting saturation
than the others. This is consistent with previous numerical studies showing that the variation of saturation with Ca depends on the viscosity ratio, fractional flow and other flow parameters \cite{kh02}. We then
identify the non-wetting clusters using Hoshen-Kopelman algorithm
\cite{hk76}. As every link can be occupied with both the fluids, a
clip threshold in the link saturation is considered to identify the
clusters \cite{trh09}. If a neighboring link has a non-wetting
saturation higher than the clip threshold, the link is then considered
to belong to the same cluster. The clusters are shown in (d), (e) and
(f) of Figures \ref{clspic_bp} and \ref{clspic_ob} for the three steady-states, where
each cluster is drawn in a different color, chosen randomly. The
distribution of the clusters shows a clear characteristic difference
of $\rm ss_1$ and $\rm ss_3$ from $\rm ss_2$ both in BP and OB. The
cluster sizes in $\rm ss_1$ and $\rm ss_3$ look very similar, and they
are distinctly different from that of $\rm ss_2$.

\begin{figure}
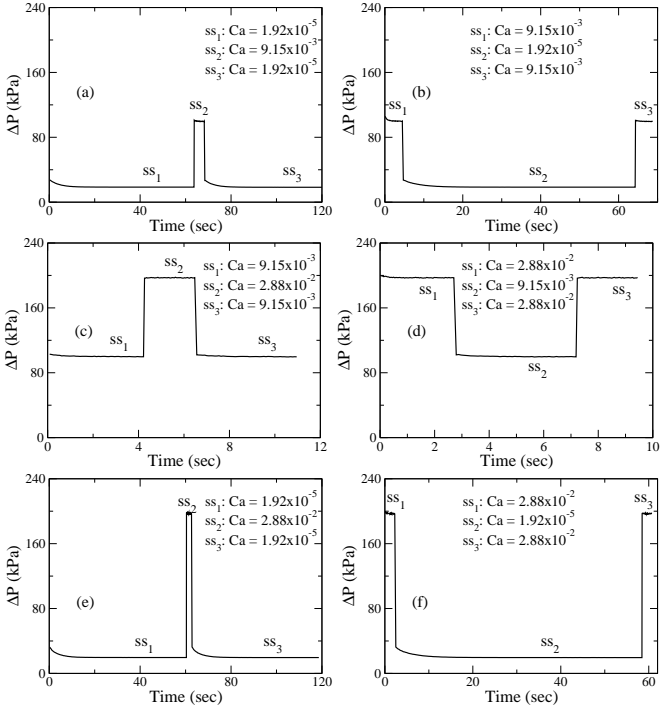

\centerline{
\includegraphics[scale=0.24,clip]{./fig11a.eps}
\includegraphics[scale=0.24,clip]{./fig11b.eps}
}
\centerline{
\includegraphics[scale=0.24,clip]{./fig11c.eps}
\includegraphics[scale=0.24,clip]{./fig11d.eps}
}
\centerline{
\includegraphics[scale=0.24,clip]{./fig11e.eps}
\includegraphics[scale=0.24,clip]{./fig11f.eps}
}
\caption{\label{pr_bp} Global pressure drop $\Delta P$ as a function
  of time for BP in different simulations with $\rm M=1$ and $S_{\rm nw}=0.74$. A rapid change in $\Delta
  P$ can be observed as soon as the overall flow rate is altered. When
  the flow rate is restored to the initial value, $\Delta P$ settle
  back to the initial steady-state value as seen for $\rm ss_1$ and
  $\rm ss_3$.}
\end{figure}

\begin{figure}[!b]
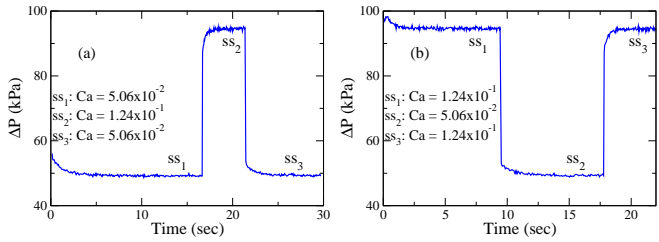

\centerline{
\includegraphics[scale=0.24,clip]{./fig12a.eps}
\includegraphics[scale=0.24,clip]{./fig12b.eps}
}
\caption{\label{pr_bpm4} Global pressure drop $\Delta P$ with time for
  BP with $\rm M=10^{-4}$ and $S_{\rm nw}=0.5$.}
\end{figure}

\begin{figure}
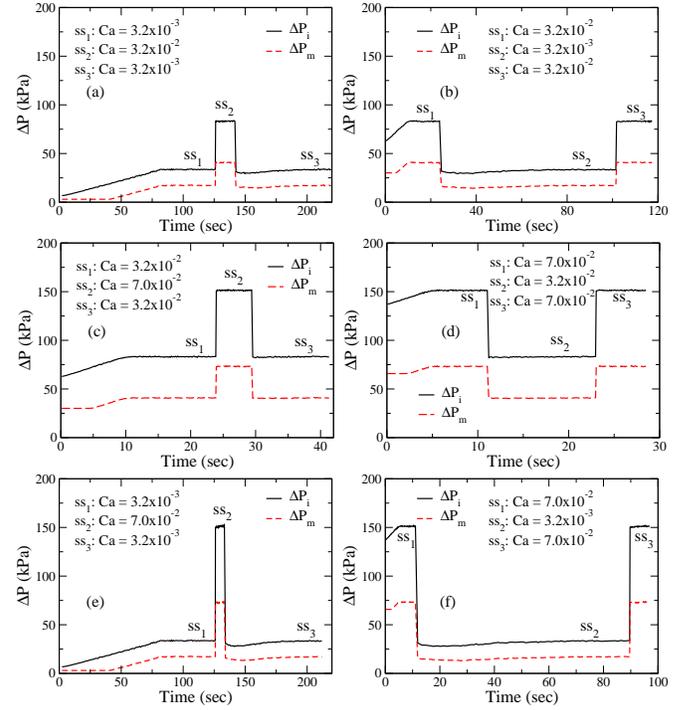

\centerline{
\includegraphics[scale=0.24,clip]{./fig13a.eps}
\includegraphics[scale=0.24,clip]{./fig13b.eps}
}
\centerline{
\includegraphics[scale=0.24,clip]{./fig13c.eps}
\includegraphics[scale=0.24,clip]{./fig13d.eps}
}
\centerline{
\includegraphics[scale=0.24,clip]{./fig13e.eps}
\includegraphics[scale=0.24,clip]{./fig13f.eps}
}
\caption{\label{pr_ob} Time evolution of global pressure drops $\Delta
  P_i$ at the inlet and $\Delta P_m$ at the middle of the system for
  OB with $\rm M=1$. A rapid change in both the pressure drops can be observed as
  soon as $\rm Ca$ is altered, but they again settle back to the
  initial value when the flow rate is restored to the initial one.}
\end{figure}

The temporal evolution of pressure drops for different simulations is
shown in Figures \ref{pr_bp} ($\rm M=1$) and \ref{pr_bpm4} ($\rm M=10^{-4}$)
for BP and in Figure \ref{pr_ob} for OB. The two different capillary
numbers for each simulation corresponding to $\rm ss_1$, $\rm ss_2$
and $\rm ss_3$ are indicated in the plots. In BP, we measure the
global pressure drop $\Delta P$ over the whole system. In OB, we
measure the pressure drops $\Delta P_i$ at the inlet nodes and $\Delta
P_m$ at the middle of the system, with respect to the outlet where the
pressures are averaged over all the nodes in the corresponding row, in
the direction perpendicular to the flow. We observe very similar
behaviors in the pressure curves as in the experiments. In the case of BP,
the system is initialized by filling the tubes with the fluids
randomly at the desired saturation $S_{\rm nw}$, which
will be constant throughout the simulation. This random initialization has the advantage of decreasing the simulation time significantly, since the steady state is reached faster than with an initial condition in which the two fluids are completely segregated. Due to this initial random
filling in BP, the global pressure $\Delta P$ starts from a higher
value and decreases with time due to the formation of
clusters. Subsequently, it reaches the steady state $\rm ss_1$ and
$\Delta P$ fluctuates around a constant average value as seen on
Figures \ref{pr_bp} and \ref{pr_bpm4}. In OB, the system is
initialized by saturating the network completely with the wetting
fluid and then the simulation is started by injecting two fluids
simultaneously through a series of alternate inlets. The flow rates of
individual fluids are controlled to obtain the required fractional flow
$F_{\rm nw}$. Both drainage and imbibition therefore take place at the
pore level creating new menisci, which increase the pressure drop
with time as seen in Figure \ref{pr_ob}. Away from the inlets, the trace of the injection channels vanishes, and a steady-state $\rm ss_1$ is attained, and
$\Delta P_i$ and $\Delta P_m$ fluctuate around constant average
values. Next, as soon as the capillary number is changed to a different
value, a rapid change in the pressure drops is observed for both BP
and OB, and the new steady-state $\rm ss_2$ is reached, characterized
by different constant values in the average pressure drops. When the
capillary number is again altered to the initial value to reach the
steady-state $\rm ss_3$, we find that the global pressure drops change back to
the initial average value. Moreover, the global pressure
drops corresponding to the same capillary numbers in different
simulations have the same average value, no matter from which
condition they have been reached.

\begin{figure}
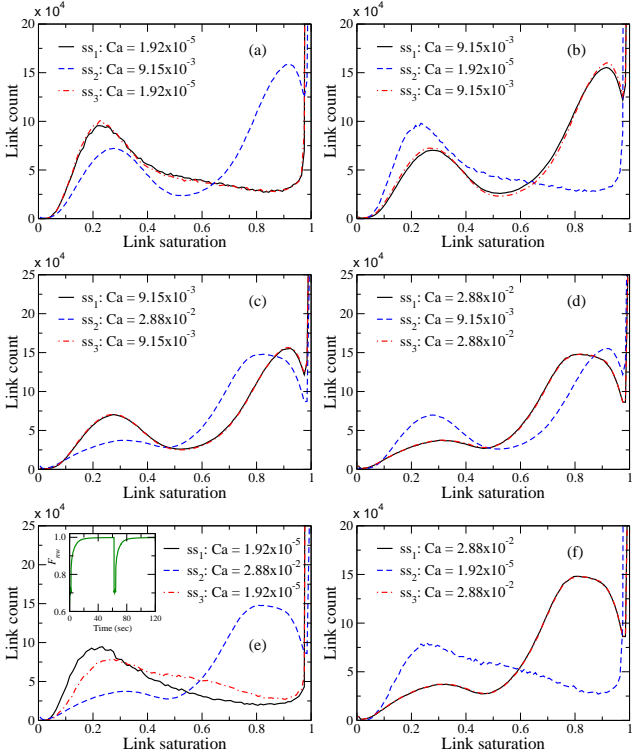

\centerline{
\includegraphics[scale=0.24,clip]{./fig14a.eps}
\includegraphics[scale=0.24,clip]{./fig14b.eps}
}
\centerline{
\includegraphics[scale=0.24,clip]{./fig14c.eps}
\includegraphics[scale=0.24,clip]{./fig14d.eps}
}
\centerline{
\includegraphics[scale=0.24,clip]{./fig14e.eps}
\includegraphics[scale=0.24,clip]{./fig14f.eps}
}
\caption{\label{hg_bp} Histograms of the network links according to
  non-wetting fluid saturation inside the links over the steady-state
  configurations for BP with $\rm M=1$ and $S_{\rm nw}=0.74$. In all the simulations, histograms are found
  similar for $\rm ss_1$ and $\rm ss_3$. Two distinct peaks are
  observed for $\rm Ca = 9.15\times 10^{-3}$, implying that the links
  are either highly saturated with the non-wetting fluid or the
  wetting fluid. The inset of (e) shows non-wetting fractional flow
  $F_{\rm nw}$ for the corresponding simulation where $F_{\rm nw} \approx
  0.99$ at $\rm Ca = 1.92\time 10^{-5}$, which is reason for minor
  variation in the histogram patterns for $\rm ss_1$ and $\rm ss_3$.}
\end{figure}

\begin{figure}[!b]
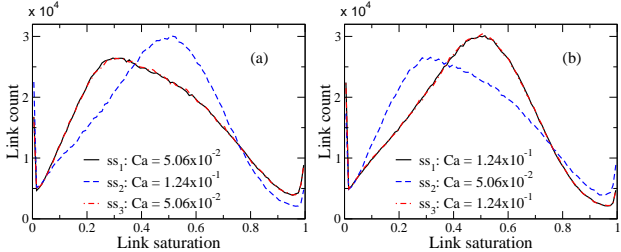

\centerline{
\includegraphics[scale=0.24,clip]{./fig15a.eps}
\includegraphics[scale=0.24,clip]{./fig15b.eps}
}
\caption{\label{hg_bpm4} Steady-state saturation histograms for BP
  with $\rm M=10^{-4}$ and $S_{\rm nw}=0.5$.}
\end{figure}

\begin{figure}
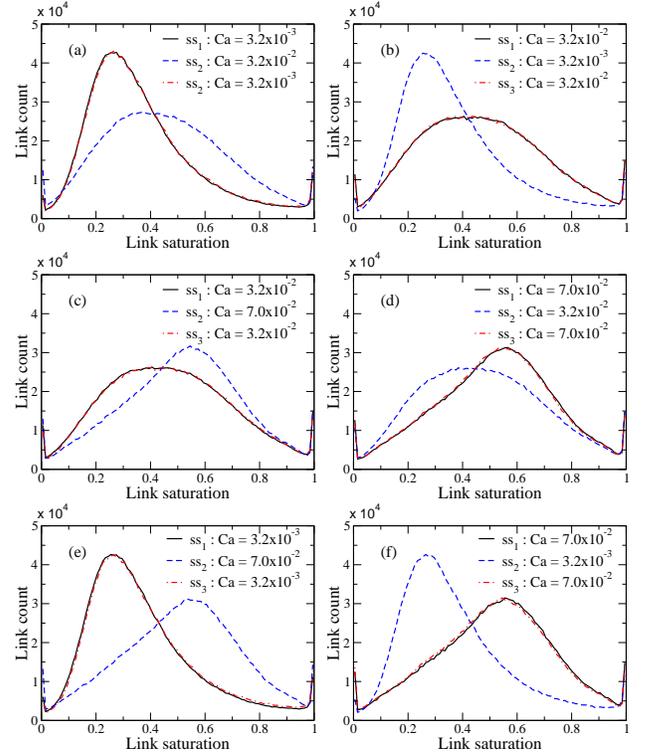

\centerline{
\includegraphics[scale=0.24,clip]{./fig16a.eps}
\includegraphics[scale=0.24,clip]{./fig16b.eps}
}
\centerline{
\includegraphics[scale=0.24,clip]{./fig16c.eps}
\includegraphics[scale=0.24,clip]{./fig16d.eps}
}
\centerline{
\includegraphics[scale=0.24,clip]{./fig16e.eps}
\includegraphics[scale=0.24,clip]{./fig16f.eps}
}
\caption{\label{hg_ob} Histograms of the links according to the
  non-wetting fluid saturation inside the links over the steady-state
  configurations for OB with $\rm M=1$. In all the simulations, histograms are found
  identical for $\rm ss_1$ and $\rm ss_3$. Histograms contain only one
  peak around the middle, which means that the links are mostly
  saturated with the mixture of both the fluids.}
\end{figure}

\begin{figure}
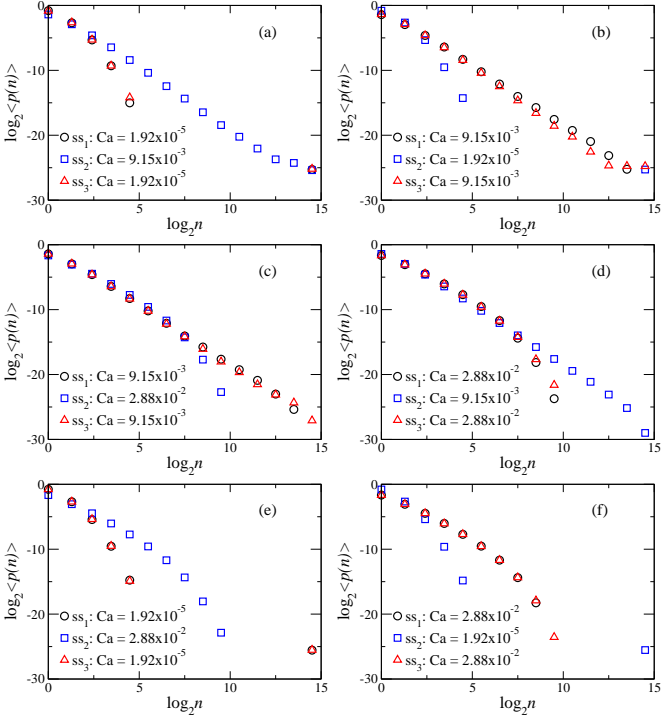

\centerline{
\includegraphics[scale=0.24,clip]{./fig17a.eps}
\includegraphics[scale=0.24,clip]{./fig17b.eps}
}
\centerline{
\includegraphics[scale=0.24,clip]{./fig17c.eps}
\includegraphics[scale=0.24,clip]{./fig17d.eps}
}
\centerline{
\includegraphics[scale=0.24,clip]{./fig17e.eps}
\includegraphics[scale=0.24,clip]{./fig17f.eps}
}
\caption{\label{cd_bp} Steady-state non-wetting cluster size
  distributions, $\langle p(n) \rangle$ \textit{vs.} $n$ for BP with $\rm M=1$ and $S_{\rm nw}=0.74$. For all the
  simulations, $\rm ss_1$ and $\rm ss_3$ are found to have the same
  distributions, which depend only on the capillary number.}
\end{figure}

\begin{figure}[!b]
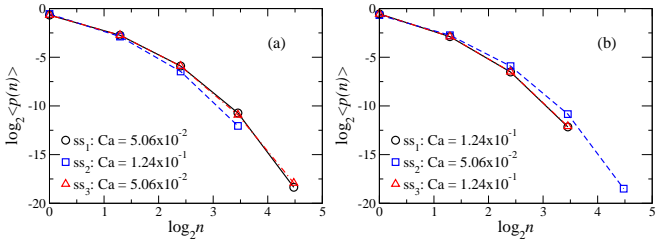

\centerline{
\includegraphics[scale=0.24,clip]{./fig18a.eps}
\includegraphics[scale=0.24,clip]{./fig18b.eps}
}
\caption{\label{cd_bpm4} Steady-state non-wetting cluster size
  distributions, $\langle p(n) \rangle$ \textit{vs.} $n$ for BP with
  $\rm M=10^{-4}$ and $S_{\rm nw}=0.5$.}
\end{figure}

\begin{figure}[!t]
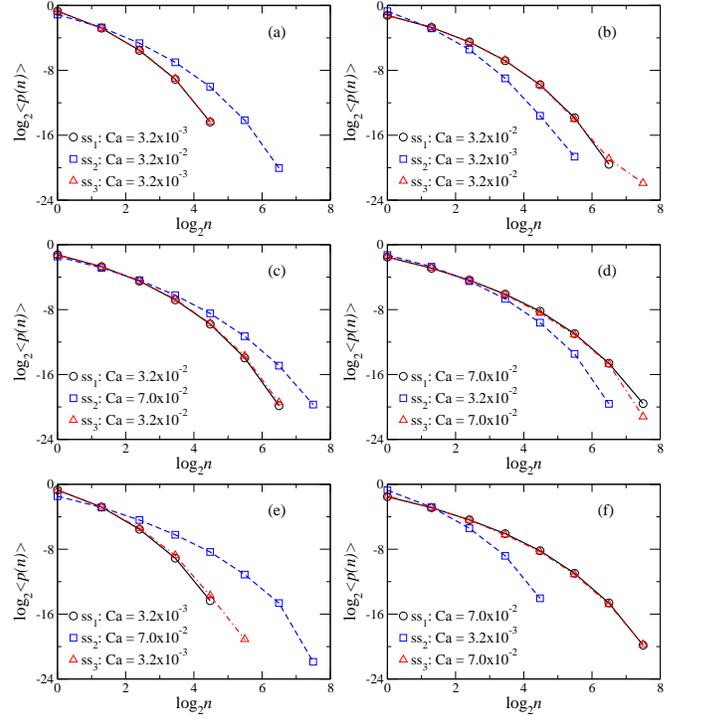

\centerline{
\includegraphics[scale=0.24,clip]{./fig19a.eps}
\includegraphics[scale=0.24,clip]{./fig19b.eps}
}
\centerline{
\includegraphics[scale=0.24,clip]{./fig19c.eps}
\includegraphics[scale=0.24,clip]{./fig19d.eps}
}
\centerline{
\includegraphics[scale=0.24,clip]{./fig19e.eps}
\includegraphics[scale=0.24,clip]{./fig19f.eps}
}
\caption{\label{cd_ob} Steady-state non-wetting cluster size
  distributions, $\langle p(n) \rangle$ \textit{vs.} $n$ for OB with $\rm M=1$. Similar to BP,
  the distributions are found identical in $\rm ss_1$ and $\rm ss_3$
  in all the simulations.}
\end{figure}

The global pressure estimates therefore completely support the
experimental observations, \textit{i.e,} that the steady state only depends on the
imposed flow rate, and not on the initial condition. Now, in order to find a
detailed microscopic information in this regard, we measure the
distribution of link saturation over the system. This measurement provides us with similar information as that of experimental gray-scale image histograms, despite being computed slighly differently. More precisely, in the experiment, the gray-scale of each
pixel is counted, where one pixel corresponds to any of the three
components -- the viscous fluid, air or the glass beads. In the
simulation, on the other hand, we count the non-wetting saturation inside each link and
compute the histogram of the link counts. Therefore, one should not try
to make a direct match of the histogram patterns from the experiments
to the simulations. The histograms in the three steady-states $\rm
ss_1$, $\rm ss_2$ and $\rm ss_3$ in different simulations are plotted
on Figures \ref{hg_bp} ($\rm M=1$) and \ref{hg_bpm4} ($\rm M=10^{-4}$) for
BP and \ref{hg_ob} for OB. In each simulation, it is very clear that
the histograms for $\rm ss_1$ and $\rm ss_3$ fall on each other
whereas they are distinctly different from that of $\rm
ss_2$. Moreover, the histogram patterns corresponding to the same Ca
in different simulations are identical. A minor difference in
the histograms for $\rm ss_1$ and $\rm ss_3$ is observed only for $\rm Ca = 1.92\times 10^{-5}$ in BP. This is due to the appearance, at low $\rm Ca$ and high
saturation in BP, of percolating non-wetting flow channels yielding a high
non-wetting fractional flow ($F_{\rm nw}\approx 0.99$), as shown in the
inset of Figure \ref{hg_bp} (e), while the rest of the system is immobilized.

 On the histograms, we observe
two distinct peaks in BP for $\rm Ca = 9.15\times
10^{-3}$: one at $s_{\rm nw} \gtrsim 0.8$ corresponding to the links
mostly filled with non-wetting fluid, and the other at $s_{\rm nw}
\lesssim 0.4$ corresponding to the links mostly filled with
wetting fluid. Therefore links can be divided into two categories:
either highly saturated with the non-wetting fluid or with the wetting
fluid, rather than containing a mixing of two phases. This in turn indicates the presence of large clusters at this $\rm Ca$, as already
observed in the fluid morphology on Figures \ref{clspic_bp} (d) and (f). For the other two Ca values in
BP, the histograms are changing towards having one peak, with a more
flat shape. For $\rm M=10^{-4}$ in
BP, and in OB, the histograms also display one peak, roughly centered but whose position shifts along the
x-axis with Ca. This indicates that most of the links are filled with
a mixture of both fluids. Therefore it is difficult to obtain
large clusters in such conditions, as observed in the fluid morphology for
OB (Figure \ref{clspic_ob}).

Finally we compute the non-wetting cluster size distributions at different
steady states. Here, the
size $n$ of a cluster is defined by the total number of links that
belong to the cluster. The probability $p(n)$ to have a $n$-sized
cluster is then defined as $p(n)=N(n)/N_{tot}$, where $N(n)$ is the
number of $n$-sized clusters out of total $N_{tot}$ clusters
identified. $p(n)$ is averaged over different configurations in the
steady state and different samples of the network. In Figures
\ref{cd_bp} and \ref{cd_ob}, $\langle p(n) \rangle$ is plotted in
log-log scale for BP and OB respectively, for different
simulations. For all the simulations, we find that the cluster size
distributions are identical for $\rm ss_1$ and $\rm ss_3$ whereas they
are different from $\rm ss_2$. Distributions for the same $\rm Ca$ for
different simulations are also the same, showing that the steady-state
cluster size distributions are history independent.

\section{Conclusions}\label{conclusions}
In this article we have considered the question of history dependence
in the steady-state two-phase flow in porous media and presented
detailed experimental and numerical investigations in this
context. Experimentally, a quasi two-dimensional laboratory model
consisting a Hele-Shaw cell filled with glass beads is considered,
through which two phases, a gas-liquid pair with a viscosity ratio
$10^{-4}$ flows at constant flow rate. The system is allowed to
evolve to a steady-state where the global pressure drop fluctuates
around a constant average value. Steady-states corresponding to the
same control parameters (e.g. capillary number) are attained from
different initial conditions. In order to characterize the complex
flow patterns in the steady-state, the gray-scale histogram of
snapshots and the non-wetting cluster size distributions are
analyzed. The system is then modeled numerically by a network of
disordered pores transporting two immiscible fluids. Steady-state
situation in the network model is attained implementing two different
boundary conditions, the toroidal one with constant saturation and the
open boundary with constant fractional flow, similar to the
experiments. The global pressure drop, distribution of non-wetting
pore saturation over the network and the cluster size distributions
are computed. Both the experimental and numerical results show that
when both the fluids are flowing in the steady state, different
measurements corresponding to the same control parameters are
identical, no matter how the steady-state has been reached. Thus,
unlike the transients, the steady-states only depend on the external
parameters, but do not depend on the initial preparation of the
system or the history of the process. We therefore conclude that, within the range of parameters explored in this study, the
steady states in simultaneous flow of two phases through porous medium
are history independent.

\begin{acknowledgments}
This work was supported by the Research Council of Norway (NFR) through PETROMAKS project nr. 193298 and CLIMIT project nr. 200041. We thank Signe Kjelstrup, Dick Bedeaux and Mihailo Jankov for fruitful discussions.
 
\end{acknowledgments}


\begin{thebibliography}{100}

\bibitem{ap95a} D.\ G.\ Avraam and A.\ C.\ Payatakes, J. Fluid
  Mech. {\bf 293}, 207 (1995).

\bibitem{bonnet77} J.\ Bonnet and R.\ Lenormand,
  Rev. Inst. Fr. Pet. {\bf 32}, 477 (1977).

\bibitem{mfj85} K.\ J.\ M{\aa}l{\o}y, J.\ Feder and T.\ J{\o}ssang,
  Phys. Rev. Lett. {\bf 55}, 2688 (1985).

\bibitem{fmsh97} O.\ I.\ Frette, K.\ J.\ M{\aa}l{\o}y, J.\ Schmittbuhl
  and A.\ Hansen, Phys. Rev. E {\bf 55}, 2969 (1997).

\bibitem{berejnov08} V.\ Berejnov, N.\ Djilali and D.\ Sinton, Lab
  Chip {\bf 8}, 689 (2008).

\bibitem{koplik85} J.\ Koplik and T.\ J.\ Lasseter,
  Soc. Petrol. Eng. J. {\bf 25}, 89 (1985).

\bibitem{lenormand88} R.\ Lenormand, E.\ Touboul and C. Zarcone {\bf
  189}, 165 (1988).

\bibitem{blunt90} M.\ Blunt and P.\ King, Phys. Rev. A {\bf 42}, 4780
  (1990).

\bibitem{cp91} G.\ N.\ Constantinides and A.\ C.\ Payatakes,
  J. Colloid Interface Sci. {\bf 141}, 486 (1991).

\bibitem{cp96} G.\ N.\ Constantinides and A.\ C.\ Payatakes, AIChE
  Journal {\bf 42}, 369 (1996).

\bibitem{rothman90} D.\ H.\ Rothman, J. Geophys. Res. {\bf 95}, 8663
  (1990).

\bibitem{rothman91} A.\ K.\ Gunstensen, D.\ H.\ Rothman, S.\ Zaleski
  and G.\ Zanetti, Phys. Rev. A {\bf 43}, 4320 (1991).

\bibitem{rothman93} A.\ K.\ Gunstensen and D.\ H.\ Rothman,
  J. Geophys. Res. {\bf 98}, 6431 (1993).

\bibitem{rothman95} B.\ Ferr\'eol and D.\ H.\ Rothman, Transp. Porous
  Media {\bf 20}, 3 (1995).

\bibitem{liu12} H.\ Liu, A.\ J.\ Valocchi and Q.\ Kang, Phys. Rev. E
  {\bf 85}, 046309 (2012).

\bibitem{aursjo2010} O.\ Aursj{\o}, H.\ A.\ Knudsen, E.\ G.\ Flekk{\o}y and K.\ J.\ M{\aa}l{\o}y, Phys. Rev. E {\bf 82}, 026305 (2010).

\bibitem{aursjo2011} O.\ Aursj{\o}, G.\ L{\o}voll, H.\ A.\ Knudsen, E.\ G.\ Flekk{\o}y and K.\ J.\ M{\aa}l{\o}y, Transport in Porous Media {\bf 86} (1), 125 (2011).

\bibitem{ww83} D.\ Wilkinson and J.\ F.\ Willemsen, J. Phys. A {\bf
  16}, 3365 (1983).

\bibitem{ws81} T.\ A.\ Witten and L.\ M.\ Sander,
  Phys. Rev. Lett. {\bf 47}, 1400 (1981).

\bibitem{paterson84} L.\ Paterson, Phys. Rev. Lett. {\bf 52}, 1621
  (1984).

\bibitem{binning99} P.\ Binning and M.\ A.\ Celia, Adv. Water
  Resour. {\bf 22}, 461 (1999).

\bibitem{ap95b} D.\ G.\ Avraam and A.\ C.\ Payatakes, Transport in
  Porous Media {\bf 20}, 135 (1995).

\bibitem{ap99} D.\ G.\ Avraam and A.\ C.\ Payatakes,
  Ind. Eng. Chem. Res. {\bf 38}, 778 (1999).

\bibitem{tap07} C.\ D.\ Tsakiroglou, D.\ G.\ Avraam and
  A.\ C.\ Payatakes, Advances in Water Resources {\bf 30}, 1981
  (2007).

\bibitem{vcp98} M.\ S.\ Valavanides, G.\ N.\ Constantinides and
  A.\ C.\ Payatakes, Transport in Porous Media {\bf 30}, 267 (1998).

\bibitem{kh02} H.\ A.\ Knudsen and A.\ Hansen, Phys.\ Rev.\ E {\bf
  65}, 056310 (2002).

\bibitem{tallakstad09} K.\ T.\ Tallakstad, G.\ L{\o}voll,
  H.\ A.\ Knudsen, T.\ Ramstad, E.\ G.\ Flekk{\o}y and
  K.\ J.\ M{\aa}l{\o}y, Phys. Rev. E {\bf 80}, 036308 (2009).

\bibitem{tkrlmtf09} K.\ T.\ Tallakstad, H.\ A.\ Knudsen, T.\ Ramstad,
  G.\ L{\o}voll, K.\ J.\ M{\aa}l{\o}y, R.\ Toussaint and
  E.\ G.\ Flekk{\o}y, Phys. Rev. Lett. {\bf 102}, 074502 (2009).

\bibitem{rassi2011} E.\ M.\ Rassi, S.\ L. Codd and J.\ D.\ Seymour, New Journal of Physics {\bf 13}, 015007 (2011).

\bibitem{sinha13} S.\ Sinha, A.\ Hansen, D.\ Bedeaux and
  S.\ Kjelstrup, Phys. Rev. E {\bf 87}, 025001 (2013).

\bibitem{sinha12} S.\ Sinha and A.\ Hansen, Europhys. Lett. {\bf 99},
  44004 (2012).

\bibitem{rh06} T.\ Ramstad and A.\ Hansen, Phys. Rev. E {\bf 73},
  026306 (2006).

\bibitem{hr09} A.\ Hansen and T.\ Ramstad, Computational Geosciences
  {\bf 13}, 227 (2009).

\bibitem{juanes2006} R.\ Juanes, E.\ J.\ Spiteri, F.\ M.\ Orr Jr. and
  M.\ J.\ Blunt, Water Resources Research {\bf 42}, W12418 (2006),
  \doi{10.1029/2005WR004806}.

\bibitem{aryana12} S.\ A.\ Aryana and A. R. Kovscek, Phys. Rev. E
  {\bf 86}, 066310 (2012).

\bibitem{greytak73} T.\ J.\ Greytak, R.\ T.\ Johnson, D.\ N.\ Paulson
  and J.\ C.\ Wheatley, Phys. Rev. Lett. {\bf 31}, 452 (1973).

\bibitem{watson96} J.\ Watson and D.\ S.\ Fisher, Phys. Rev. B {\bf
  54}, 938 (1996).

\bibitem{Cheng2008} N.-S.\ Cheng, Ind. Eng. Chem. Res. {\bf 47}, 3285
  (2008).

\bibitem{imageJ} W.\ S.\ Rasband, ImageJ, U.\ S.\ National Institutes of Health, Bethesda, Maryland, USA, http://imagej.nih.gov/ij/, 1997-2012

\bibitem{amhb98} E.\ Aker, K.\ J.\ M\r{a}l\o y, A.\ Hansen and
  G.\ G.\ Batrouni, Transp. Porous Media {\bf 32}, 163 (1998);
  E. Aker, K.\ J.\ M\r{a}l\o y and A.\ Hansen, Phys. Rev. E {\bf 58},
  2217 (1998).

\bibitem{d92} F.\ A.\ L.\ Dullien, {\em Porous Media: Fluid Transport
  and Pore Structure} (Academic Press, San Diego, 1992).

\bibitem{w21} E.\ W.\ Washburn, Phys. Rev. {\bf 17}, 273 (1921).

\bibitem{kah02} H.\ A.\ Knudsen, E.\ Aker and A.\ Hansen,
  Transp. Porous Media {\bf 47}, 99 (2002).

\bibitem{hk76} J.\ Hoshen and R.\ Kopelman, Phys. Rev. B {\bf 14},
  3438 (1976).

\bibitem{trh09} G.\ T{\o}r{\aa}, T.\ Ramstad and A.\ Hansen,
  Europhys. Lett. {\bf 87}, 54002 (2009).

\bibitem{sa92} D.\ Stauffer and A.\ Aharony, {\it Introduction to
  Percolation Theory} (Taylor \& Francis, London, 1992).


\end{thebibliography}
\end{document}